\documentclass[onefignum,onetabnum]{siamsials251208}

\usepackage{lipsum}
\usepackage{booktabs}
\usepackage{tikz}
\usetikzlibrary{arrows.meta, decorations.markings, calc}
\usetikzlibrary{shapes.geometric, positioning}
\usepackage{amsfonts}
\usepackage{graphicx}
\usepackage{epstopdf}
\usepackage{algorithmic}
\usepackage{subcaption}
\usepackage{amssymb}
\ifpdf
  \DeclareGraphicsExtensions{.eps,.pdf,.png,.jpg}
\else
  \DeclareGraphicsExtensions{.eps}
\fi

\usepackage{enumitem}
\setlist[enumerate]{leftmargin=.5in}
\setlist[itemize]{leftmargin=.5in}

\newsiamremark{remark}{Remark}
\newsiamremark{hypothesis}{Hypothesis}
\crefname{hypothesis}{Hypothesis}{Hypotheses}
\newsiamthm{claim}{Claim}
\newsiamremark{fact}{Fact}
\crefname{fact}{Fact}{Facts}

\headers{MFPT for Persistent random walk}{F. Saghafifar, and D. Coombs}

\title{Mean First Passage Time for Persistent Random Walks in Annular Search Domains
\thanks{\funding{This work was supported by the Natural Sciences and Engineering Research Council of Canada.}}
}

\author{Fatemeh Saghafifar\thanks{Department of Mathematics and Institute of Applied Mathematics, University of British Columbia, 1984 Mathematics Road, Vancouver, BC, V6T 1Z2, Canada
  (\email{fsaghafi@math.ubc.ca}).}
\and Daniel Coombs\thanks{Department of Mathematics and Institute of Applied Mathematics, University of British Columbia, 1984 Mathematics Road, Vancouver, BC, V6T 1Z2, Canada
  (\email{coombs@math.ubc.ca}).}}

\usepackage{amsopn}

\ifpdf
\hypersetup{
  pdftitle={Mean First Passage Time for persistent random walks in annular search domains},
  pdfauthor={Fatemeh Saghafifar, Daniel Coombs}
}
\fi

\usepackage{xspace}

\usepackage{enumitem}

\begin{document}

\maketitle

% REQUIRED
\begin{abstract}
We study the mean first-passage time of a random walker to a small absorbing target at the center of a two-dimensional annulus with a specularly reflecting outer boundary. The problem is motivated by natural killer cell migration toward a target cancer cell, where the goal is to quantify how long it takes immune cells to reach the target and how search efficiency depends on directional persistence and chemotactic bias. Cell motion is modeled as a velocity-jump process. We first consider a correlated random walk with a von Mises turning kernel, with a concentration parameter controlling directional persistence. We then extend the model to a biased correlated random walk using a phase-shifted turning kernel that represents preferential motion, for example following a concentration gradient. Our analysis combines closed-form benchmarks for simple and biased random walks, Fourier-mode reductions of the transport equations for the correlated and biased correlated models, and a fast-turning perturbation expansion that gives an analytical correction to the diffusion-limit mean first-passage time for the random walker. Our analytical results are supported by numerical methods that include a semi-Lagrangian solver in radial and angular coordinates, a stationary discretisation designed to handle biased transport, and an event-driven Monte Carlo simulator for cross-validation. Together, our results provide a framework relating persistent and biased immune-cell motion to target-search times in confined two-dimensional domains.
\end{abstract}

% REQUIRED

% REQUIRED

\section{Introduction}
Many cell movements exhibit directional persistence: bacteria run and
tumble \cite{BergBrown1972,Berg2004}, T~cells and neutrophils alternate between smooth runs with sudden turns \cite{MillerEtAl2002,
JannatEtAl2010}, fibroblasts and \emph{Dictyostelium} retain memory of
their direction over minutes \cite{GailBoone1970,VanHaastertBosgraaf2009},
and keratocytes provide an example of highly persistent cell migration with trajectories that remain close to ballistic \cite{EuteneuerSchliwa1984,LeeEtAl1993,KerenEtAl2008}.
One biologically important quantity in all these scenarios is the first-passage time to a target, such as an antigen-presenting cell, a chemical source, or a wound edge, under stochastic cell-migration dynamics.

The usual random walk models, simple random walks (SRWs) and biased random walks (BRWs), are widely used in mathematical biology to describe single-cell tracking data and can yield closed-form mean first passage time (MFPT) formulae in radially symmetric geometries \cite{plank2025random,OthmerHillen2002,Pavliotis2014}. However, these models do not distinguish between two different sources of directed motion: directional persistence, which arises from a walker's internal memory of its previous direction, and external bias, which arises from cues such as chemical gradients. This distinction is important because the underlying stochastic process, and therefore the resulting search dynamics, depends on how directional information is introduced. The correlated random walk (CRW) accounts for directional persistence by sampling turning angles from a non-uniform distribution, typically the von Mises distribution \cite{MardiaJupp2000,CodlingPlankBenhamou2008}. The corresponding backward equation is a transport equation in position-velocity space,
rather than a diffusion equation in postion space alone, and it admits no closed-form MFPT in confined geometries.

Related work on MFPT for non-diffusive motion has been developed along two
main lines: For velocity-jump and active-particle models, recent
results include a general transport-MFPT framework by Hillen,
D'Orsogna, Mantooth, and Lindsay~\cite{hillen2025mean}, its recent 
extension to higher
dimensions with directional bias by D'Orsogna, Lindsay, and
Hillen~\cite{d2026mean}, and a MFPT
analysis for active Brownian particles in two dimensions by Iyaniwura
and Peng~\cite{iyaniwura2025mean}. 
Earlier work in this direction includes the quasi-steady-state
analysis of motor-driven transport on two-dimensional microtubular
networks by Bressloff and Newby~\cite{bressloff2011quasi}, and the
asymptotic narrow-capture analysis of first-passage problems in
ecological settings by Kurella, Tzou, Coombs, and
Ward~\cite{kurella2015asymptotic}. For diffusive
motion in domains with small targets, the narrow-capture problem has an
extensive literature; for reviews see Holcman and Schuss~\cite{holcman2014}
and the book by Grebenkov, Metzler, and Oshanin~\cite{grebenkov2024book}.
This literature covers the MFPT and its full distribution, including the
variance~\cite{grebenkov2019full}, multi-target configurations with mixed
boundary conditions~\cite{grebenkov2026competition}, trap-placement
optimization~\cite{cheviakov2024optimization}, and the extreme first-passage
time of the fastest among multiple independent walkers competing to reach a
target~\cite{lawley2020universal,grebenkov2020fastest}. 

%The present paper addresses 
Here, we address a complementary regime: rather than studying diffusive searchers in domains with small absorbing windows, we consider a velocity-jump process in which persistence and chemotactic bias are introduced separately and their combined effect on MFPT is quantified through the correlated and biased-correlated random walk pair. To focus on the main issues, we study a
simplified but representative geometry: the two-dimensional annulus
$a < |\mathbf{x}| < R$, with an absorbing inner boundary and a specularly
reflecting outer boundary.

In this setting, we present several several new results relevant to biological modelling:
\begin{enumerate}[label=(\roman*)]
    \item We derive analytical limits
of the CRW MFPT in the fast turning and strong persistence regimes,
and compute a closed-form expression for the
persistence-induced reduction in search time via a leading-order perturbation correction to the diffusion-limit
MFPT. This correction arises from a kinetic boundary layer at the absorbing target, rather than from a bulk correction in the interior. 
\item We present a chemotactic extension model in which
persistence and external bias act together, represented here by a biased
correlated random walk (BCRW).
\item We develop a semi-Lagrangian numerical
solver for the CRW MFPT in confined geometries which resolves a
persistence-induced outer wall layer. We characterize the solver's structural accuracy
ceiling through eigenpair analysis and Richardson extrapolation, and replace
it with a stationary formulation that eliminates this ceiling. We then extend the solver to the BCRW and validate against an event-driven Monte Carlo simulator and ballistic regime predictions from the ecological random walk
literature~\cite{CodlingThesis2003,CodlingPlankBenhamou2008}. 
\end{enumerate}

Together, our results show that
directional persistence reduces the confined search time monotonically over
the biologically relevant parameter range, with diminishing additional
benefit at high persistence.

The paper is organized as follows. \Cref{sec:formulation} sets up the
geometry, the backward Fokker--Planck equation, and the CRW and BCRW
transport equations. 
\Cref{sec:analysis} summarizes a number of useful analytical results: a double-integral MFPT formula for the SRW and BRW which serves as a useful benchmark; fast-turning and strong-persistence
limits of the CRW; a Fourier-mode approach to the CRW and BCRW problems; and finally the leading-order perturbation correction to the diffusion-limit MFPT for the CRW.
\Cref{sec:numerics} describes the CRW semi-Lagrangian solver, the BCRW extension, and the Monte Carlo benchmark. 
\Cref{sec:verification} reports verification, graded-mesh convergence, eigenpair
analysis, Richardson extrapolation, and the documented ${\sim}2.7\%$
structural ceiling. 
\Cref{sec:bio} provides a rough calibration of $\kappa$ using data from different classes of cells and
reports a worked prediction for T cells. Finally, \Cref{sec:mainresult} states the main verified
result and \Cref{sec:discussion} interprets the findings and describes possible future work.

%%%%%%%%%%%%%%%%%%%%%%%%%%%%%%%%%%%%%%%
\begin{table}[]
    \centering
    \begin{tabular}{c|c}
       Abbreviation  & Model \\
       \hline
       SRW  &  Simple Random Walk (diffusive motion)\\
       BRW  &  Biased Random Walk (e.g. SRW with chemoattractants)\\
       CRW  &  Correlated Random Walk (e.g. SRW with persistent motion)\\
       BCRW &  Biased Correlated Random Walk (full model) \\
    \end{tabular}
    \caption{Abbreviated model names}
    \label{tab:my_label}
\end{table}
%%%%%%%%%%%%%%%%%%%%%%%%%%%%%%%%%%%%%%%

\section{Problem formulation}\label{sec:formulation}

\begin{figure}[t]
\centering
\begin{tikzpicture}[scale=0.82]

% ================= Left panel: general domain =================
\begin{scope}[shift={(0,0)}]

% General curved domain
\draw[thick]
(0.2,1.6) .. controls (0.7,2.4) and (2.4,2.3) .. (2.8,1.7)
.. controls (3.2,1.1) and (2.4,0.9) .. (2.8,0.2)
.. controls (3.2,-0.6) and (0.8,-0.5) .. (0.4,0.1)
.. controls (-0.1,0.8) and (-0.2,1.3) .. (0.2,1.6) -- cycle;

% Small target
\fill[green!50!black] (1.85,0.75) circle (0.05);
\draw[green!50!black, thick] (1.85,0.75) circle (0.05);

% Labels
\node at (1.45,1.25) {$D$};
\node at (1.35,-0.72) {$\partial D$};
% \node[font=\scriptsize] at (2.35,0.75) {target};

% Subcaption
\node[align=center, font=\scriptsize] at (1.45,-1.35) {(a) General confined domain};

\end{scope}

% ================= Right panel: annular domain =================
\begin{scope}[shift={(7,0.65)}]

% Outer reflecting boundary
\draw[thick] (0,0) circle (1.55);

% Inner absorbing target
\fill[green!50!black] (0,0) circle (0.16);
\draw[green!50!black, thick] (0,0) circle (0.16);

% Radius R
\draw[densely dotted, ->] (0,0) -- (-1.1,1.1);
\node at (-0.65,0.75) {$R$};

% Target diameter 2a
\draw[|-|] (-0.16,-0.42) -- (0.16,-0.42);
\node at (0,-0.68) {$2a$};

% Reflecting label and arrow
\draw[blue!70, ->, thick] (0,1.55) -- (-0.35,1.95);
\node[blue!70, font=\small] at (-0.75,2.15) {reflecting};

% Absorbing label and arrow
\draw[blue!70, ->, thick] (0.95,0.75) -- (0.18,0.14);
\node[blue!70, font=\small] at (1.35,0.85) {absorbing};

% Subcaption aligned with left
\node[align=center, font=\scriptsize] at (0,-2.00) {(b) Simplified annular domain};

\end{scope}
% ================= Panel c: absolute and relative heading =================
\begin{scope}[shift={(11,-0.1)}, scale=0.9]

% Origin and point
\coordinate (O) at (0,0);
\coordinate (P) at (2.6,1.45);

% Axes
\draw[->, gray] (-0.2,0) -- (4.2,0) node[right] {$x$};
\draw[->, gray] (0,-0.2) -- (0,3.0) node[above] {$y$};

% Radial vector
\draw[->, thick] (O) -- (P);
\node[below left] at (O) {$0$};
\node[above left] at (1.25,0.72) {$r$};
\node[below right] at ($(P)+(0.05,-0.05)$) {$(r,\theta)$};

% Local radial direction through P
\draw[dashed] (P) -- ++(1.25,0.70);

% Absolute heading reference from P
\draw[dashed, gray] (P) -- ++(1.45,0);

% Heading vector
\draw[->, thick] (P) -- ++(1.15,1.75);
\node at ($(P)+(1.35,1.95)$) {$\mathbf{v}$};

% Angle theta at origin
\draw[->] (0.62,0) arc[start angle=0,end angle=29.2,radius=0.62];
\node at (0.95,0.14) {$\theta$};

% Angle phi at P, from horizontal to heading
\draw[->, blue!70] ($(P)+(0.75,0)$) arc[start angle=0,end angle=56.7,radius=0.75];
\node[blue!70] at ($(P)+(1.25,0.30)$) {$\phi$};

% Relative heading psi, from radial direction to heading
\draw[->, green!50!black] ($(P)+(0.92,0.51)$) arc[start angle=29.2,end angle=56.7,radius=1.05];
\node[green!50!black] at ($(P)+(2,0.92)$) {$\psi=\phi-\theta$};

% Point
\fill (P) circle (1.5pt);

% Subcaption
\node[align=center, font=\scriptsize] at (2.0,-1.35) {(c) Absolute and relative headings};

\end{scope}
\end{tikzpicture}
\caption{Schematic of the search problem. 
(a) A general confined domain with an absorbing small target (green node). 
(b) The simplified annular geometry, where the inner boundary $r=a$ is absorbing and the outer boundary $r=R$ is specularly reflecting.
(c) Definition of the absolute heading $\phi$, polar angle $\theta$, and relative heading $\psi=\phi-\theta$.}
\label{fig:domains}
\end{figure}
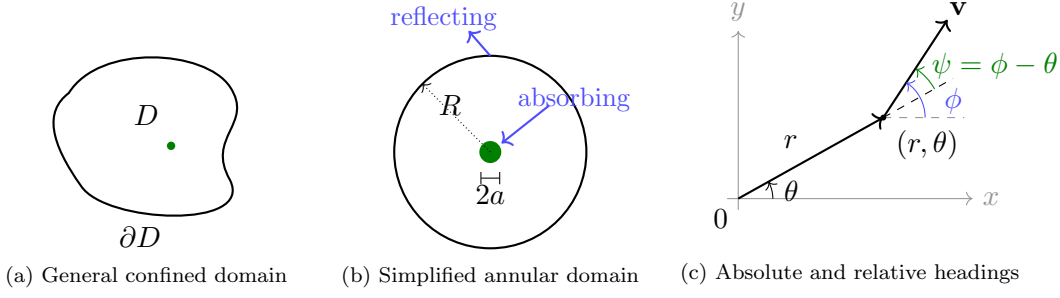

We formulate the target-search problem in a two-dimensional annular domain
\[
D=\{\,\mathbf{x}\in\mathbb{R}^2: a<|\mathbf{x}|<R\,\},
\]
where the inner boundary $r=a$ is absorbing and represents the target, such as a tumour or a cancer cell, while the outer boundary $r=R$ is reflecting and represents confinement by the surrounding tissue or an experimental domain. This geometry provides a simplified model of a confined search process in which a motile cell moves within a bounded region, and first passage occurs when the cell reaches the central target (\Cref{fig:domains}).

\paragraph{Mean First Passage Time backward formulation} We first recall the standard backward formulation for Brownian motion-based models. For a time-homogeneous stochastic differential equation
\[
dX_t=b(X_t)\,dt+\sigma(X_t)\,dW_t
\]
on a bounded domain $D\subset\mathbb{R}^2$, the infinitesimal generator acting on smooth test functions $f(X_t)$ is defined as
\[
\mathcal{L}f
=
b\cdot\nabla f
+
\left(\frac{1}{2}\sigma\sigma^\top\right):\nabla^2 f .
\]
This operator is used to derive the mean first-passage time of the stochastic process which is defined as
\[
T(\mathbf{x})
=
\mathbb{E}\!\left[
\inf\{t>0:X_t\in\partial D\}\mid X_0=\mathbf{x}
\right],
\]
and satisfies the backward Fokker--Planck equation
\begin{equation}
\mathcal{L}T(\mathbf{x})=-1,
\qquad \mathbf{x}\in D,
\label{eq:bfpe}
\end{equation}
with $T=0$ on the absorbing boundary and a homogeneous Neumann condition on any reflecting boundary \cite{Pavliotis2014}. 

\paragraph{Velocity-jump process and turning kernel} To model directional persistence, we replace the diffusion process with a velocity-jump process, which retains information about the walker’s previous direction. The walker moves at constant speed $s$ 
%$s=|\mathbf{v}|$
in heading direction $\phi$. Turning events occur as a Poisson process with rate $\mu$, and after each event the new heading is sampled from a turning kernel $K(\phi\mid\phi')$, where $\phi'$ is the heading before the turn. Between turning events, the walker follows straight-line motion at speed $s$.

Because the target and outer boundary are concentric, polar coordinates are the natural choice. The boundaries are simply $r=a$ and $r=R$, and the relative heading $\psi=\phi-\theta$ separates radial motion toward or away from the target from tangential motion around it. 

For the (unbiased) correlated random walk (CRW), the turning kernel depends
only on the turning angle $\eta = \phi - \phi'$. Since headings are angular
variables defined on the circle, the turning-angle distribution should also be
circular. We therefore use the von Mises distribution, which is the circular
analogue of the normal distribution:
\begin{equation}
\label{eq:vonmises}
K(\eta;\kappa) = \frac{1}{2\pi I_0(\kappa)}\exp(\kappa\cos\eta),
\qquad \eta = \phi - \phi'.
\end{equation}
Here $\kappa \ge 0$ is the concentration parameter, and the normalization
contains $I_n$, the modified Bessel function of the first kind of order $n$.
The parameter $\kappa$ controls the directional persistence of the walk. The walk's
first trigonometric moment is
\begin{equation*}
c = \langle\cos\eta\rangle = \frac{I_1(\kappa)}{I_0(\kappa)},
\end{equation*}
so $\kappa = 0$ gives an isotropic turning distribution, while larger $\kappa$
corresponds to stronger directional memory \cite{CodlingThesis2003, CodlingPlankBenhamou2008, MardiaJupp2000}.
Because $K$ depends only on the turning angle $\eta = \phi - \phi'$ and cosine
is an even function, the kernel is symmetric under interchange of its pre- and post-turn
headings and is therefore self-adjoint; the adjoint kernel $K^*$ has the same
functional form as~\eqref{eq:vonmises}.

In the unbiased case, rotational symmetry implies that the mean first-passage
time depends only on the radius $r$ and the relative heading $\psi = \phi -
\theta$, so we write $T = T(r,\psi)$. The deterministic motion between turning
events gives the advection part of the backward operator. Since the walker
moves at speed $s$ in the absolute heading direction $\phi$, the polar-coordinate
rates are
\begin{equation*}
\frac{\partial r}{\partial t} = s\cos\psi, \qquad
\frac{\partial \theta}{\partial t} = \frac{s}{r}\sin\psi, \qquad
\frac{\partial \phi}{\partial t} = 0,
\end{equation*}
and the relative heading satisfies
\begin{equation*}
\frac{\partial \psi}{\partial t}
= \frac{\partial \phi}{\partial t} - \frac{\partial \theta}{\partial t}
= -\frac{s}{r}\sin\psi.
\end{equation*}
Turning events occur at rate $\mu$. At a turning event, the current value of
$T(r,\psi)$ is replaced by its average over post-turn headings weighted by the
turning kernel, expressed in the relative-heading variable as
$K(\psi - \psi';\kappa)$. Combining the deterministic streaming terms, the
turning operator, and the fact that time accumulates at unit rate before
absorption, we obtain the backward transport equation
\begin{equation}
\label{eq:crw-backward}
-1 = s\cos\psi\,\partial_r T
- \frac{s\sin\psi}{r}\,\partial_\psi T
- \mu T
+ \mu\int_0^{2\pi} K(\psi - \psi';\kappa)\,T(r,\psi')\,d\psi'.
\end{equation}
The inner boundary condition is that $T(a,\psi) = 0$ for inward-pointing headings. At the outer boundary, we set
$T(R,\psi) = T(R,\pi-\psi)$. The outer condition states that, upon contact with
the reflecting wall, the radial component of velocity changes sign while the
tangential component is preserved.

\paragraph{Biased Correlated random walk and chemotaxis} To include chemotactic bias in addition to persistence, we introduce a biased correlated random walk. Motivated by biased and correlated random walk formulations in the biological movement literature \cite{CodlingThesis2003,CodlingPlankBenhamou2008,plank2025random}, we introduce bias by shifting the turning distribution toward a preferred direction $\phi_0(r,\theta)$, which represents the local direction of an external cue such as a chemical gradient produced by the target. The bias strength is denoted by $d_\tau$. A phase-shifted von Mises kernel is used:
\begin{equation}
K(\phi\mid r,\theta,\phi')
=
\frac{1}{2\pi I_0(\kappa)}
\exp\!\left[
\kappa\cos\!\left(
\phi-\phi'
+d_\tau(\phi-\phi_0(r,\theta))
\right)
\right].
\label{eq:bcrw-kernel}
\end{equation}
When $d_\tau=0$, this reduces to the unbiased correlated random walk kernel. When $d_\tau\neq 0$, the new heading is influenced both by the previous heading, through persistence, and by the external bias direction, through the phase shift.

Because the preferred direction may depend on position, the biased model is no longer rotationally invariant in the same reduced variables unless the biasing (cue) field has special symmetry. We therefore write the mean first-passage time as $T=T(r,\theta,\phi)$. The backward transport equation for the biased correlated random walk is
\begin{equation}
-1
=
s\cos(\phi-\theta)\,\partial_r T
+
\frac{s\sin(\phi-\theta)}{r}\,\partial_\theta T
-\mu T
+\mu\int_0^{2\pi}
K^*(\phi,\phi',r,\theta)\,
T(r,\theta,\phi')\,d\phi',
\label{eq:bcrw-bte}
\end{equation}
where the adjoint kernel is obtained by swapping the pre-turn and post-turn headings,
\[
K^*(\phi,\phi',r,\theta)
=
K(\phi'\mid r,\theta,\phi).
\]
The absorbing condition at the target is
\begin{equation}
T(a,\theta,\phi)=0
\qquad
\text{for headings that enter the target},
\label{eq:bcrw-abs}
\end{equation}
and the specular reflection condition at the outer boundary, based on calculations shown in \Cref{app:specular}, will be
\begin{equation}
T(R,\theta,\phi)
=
T(R,\theta,2\theta+\pi-\phi).
\label{eq:bcrw-reflect}
\end{equation}
This condition is the angular form of specular reflection: the outgoing heading is obtained by reflecting the incoming velocity across the tangent line at the boundary.

\paragraph{Model hierarchy} The simple random walk, biased random walk, correlated random walk, and biased correlated random walk therefore form a hierarchy of models. The simple and biased random walks provide diffusion-based benchmark problems for which closed-form radial mean first-passage times can be derived. The correlated random walk introduces directional memory and leads to a transport equation in position-heading space. The biased correlated random walk further includes external directional cues and provides the framework used below to study the combined effect of persistence and chemotactic bias on search time.
%%%%%%%%%%%%%%%%%%%%%%%%%%%%%%%%%%%%%%%%%%%%%%%%%%
\section{Analytical results}\label{sec:analysis}
\subsection{Simple and Biased Random Walk}\label{sec:analysis-srwbrw}

\paragraph{A unified double-integral formula}
We begin with a simple diffusion-based model. For the constant diffusivity simple random walk, with stochastic differential equation written as
\[
dX_t=\sqrt{2D}\,dW_t,
\]
the backward Fokker--Planck equation \eqref{eq:bfpe} reduces, under radial symmetry, to
\[
D\left(u''(r)+\frac{1}{r}u'(r)\right)=-1,
\qquad
u(a)=0,
\qquad
u'(R)=0.
\]
Solving this boundary value problem gives
\begin{equation}
u_{\mathrm{SRW}}(r)
=
\frac{R^2}{2D}\ln\frac{r}{a}
+
\frac{a^2-r^2}{4D}.
\label{eq:srw}
\end{equation}

The same calculation extends to a biased random walk with radial drift and possibly anisotropic diffusivities. If the radial drift is written as $-\mu(r)\mathbf{e}_r$ and the radial and angular diffusivities are $D_r(r)$ and $D_\theta(r)$, respectively, then the radial mean first passage time satisfies
\[
D_r(r)u''(r)
+
\left(\frac{D_\theta(r)}{r}-\mu(r)\right)u'(r)
=
-1.
\]
Writing $w=u'$ gives a first-order linear equation for $w$. Defining
\[
P(r)=\frac{D_\theta(r)/r-\mu(r)}{D_r(r)},
\qquad
M(r)=\exp\!\left(\int^r P(q)\,dq\right),
\]
and using the reflecting condition $u'(R)=0$, we obtain
\begin{equation}
u_{\mathrm{BRW}}(r)
=
\int_a^r
\frac{1}{M(t)}
\int_t^R
\frac{M(s)}{D_r(s)}
\,ds\,dt .
\label{eq:brw-double-integral}
\end{equation}
This double-integral representation provides a common benchmark for the constant and variable coefficient simple and biased random walk cases used in this paper. In particular, the constant drift, isotropic diffusion biased random walk is recovered as a special case and can also be written in terms of exponential integral functions. 

Numerical finite difference methods for the simple diffusion-based walks (SRW and BRW) are described in \Cref{app:fd}.

\paragraph{Fast-turning diffusion limit}
We next consider the fast-turning limit of the correlated random walk. In this regime, turning events occur frequently compared with the time scale of target search, so the transport process can be approximated by an effective diffusion. A moment closure of the forward CRW equation gives the following macroscopic density equation \cite{hillen2025mean, HillenOthmer2000}.
\begin{equation}
\partial_t\rho
=
D_{\mathrm{eff}}\Delta\rho,
\quad \text{where} \quad
D_{\mathrm{eff}}
=
\frac{s^2}{2\mu(1-m_1)}
\quad \text{and} \quad 
m_1=\frac{I_1(\kappa)}{I_0(\kappa)} .
\label{eq:Deff}
\end{equation}
The same diffusion-limit equation arises as the unbiased, fast-turning limit of the general higher-dimensional velocity-jump MFPT framework
of reference \cite{d2026mean}, in the radially symmetric unbiased case.
The leading-order mean first passage time is therefore the SRW \eqref{eq:srw} with $D=D_{\mathrm{eff}}$. This diffusion approximation provides the reference solution used below to measure the effect of finite turning rate and directional persistence on the CRW MFPT.

\paragraph{Strong-persistence limit}\label{sec:analysis-strong}
We also record the limiting form of the correlated random walk when directional persistence is strong. As the concentration parameter $\kappa$ tends to infinity, the von Mises kernel concentrates near zero turning angle and is locally Gaussian with variance $1/\kappa$. In this limit, the turning operator is approximated by angular diffusion, so that
\[
\mu\int_0^{2\pi}K^*(\psi,\psi')T(r,\psi')\,d\psi' - \mu T(r,\psi)
\;\approx\;
D_\psi\,\partial_{\psi\psi}T,
\quad \text{where} \quad
D_\psi=\frac{\mu}{\kappa}.
\]
Thus \Cref{eq:crw-backward} approaches a drift--diffusion equation in the heading variable, closely related to the backward equations used for active Brownian particles \cite{iyaniwura2025mean}. The leading-order motion is nearly deterministic between rare changes in heading, so the MFPT approaches the straight-line hitting time when the initial orientation intersects the target. In the annular problem, this limit is modified by specular reflection at the
outer boundary. The near-ballistic regime also develops a thin boundary layer
in heading space, which is a common feature of strongly persistent velocity-jump
processes. This is also the regime in which our semi-Lagrangian numerical solver begins to lose angular resolution (scenarios presented in \Cref{sec:verification}).

\paragraph{Reduction of the CRW Fourier hierarchy}
We next reduce the unbiased transport equation by using the periodicity of the heading variable. Since the von Mises kernel depends only on the turning angle, it acts as a convolution operator on the circle. Expanding
\[
T(r,\psi)=\sum_{n\in\mathbb Z}u_n(r)\mathrm{e}^{in\psi},
\]
the convolution is diagonal in Fourier space, with multipliers given by the normalized moments of the von Mises distribution,
\begin{equation}
(K*T)(\psi)
=
\sum_{n\in\mathbb Z}
m_n u_n(r)\mathrm{e}^{in\psi},
\qquad
m_n=\frac{I_n(\kappa)}{I_0(\kappa)} .
\label{eq:K-fourier-crw}
\end{equation}
This diagonalization is the circular analogue of using Fourier modes for translation invariant kernels and follows from the standard Fourier representation of the von Mises distribution \cite{MardiaJupp2000,Hillen2017VonMises}.

Substituting this expansion into \Cref{eq:crw-backward} and matching coefficients of $\mathrm{e}^{in\psi}$ gives the infinite coupled hierarchy
\begin{equation}
-\delta_{n,0}
=
\frac{s}{2}\bigl(u'_{n-1}+u'_{n+1}\bigr)
+\frac{s}{2r}\bigl((n-1)u_{n-1}+(n+1)u_{n+1}\bigr)
+\mu\bigl(m_n-1\bigr)u_n .
\label{eq:crw-fourier-hierarchy}
\end{equation}
The coupling between neighboring modes comes from the factors $\cos\psi$ and $\sin\psi$ in the advection part of the transport equation, while the turning operator remains diagonal in mode space. The absorbing condition gives $u_n(a)=0$ for all $n$, and specular reflection at the outer boundary gives
\[
u_n(R)=(-1)^n u_{-n}(R).
\]

This hierarchy is useful analytically because it makes explicit how directional persistence enters through the multipliers $m_n$. It also provides a diagnostic check on the numerical transport solver. In this work, however, truncated-Fourier versions of \Cref{eq:crw-fourier-hierarchy} are not used as the primary solver. Direct truncations led to structural difficulties that are summarized in \Cref{app:dae}. We therefore use the hierarchy for analysis and verification, while the main computations are carried out with the stationary transport discretization described in \Cref{sec:numerics}.

\paragraph{Reduction of the BCRW Fourier hierarchy}\label{sec:analysis-fourier-bcrw}
We next explore how the biased turning operator appears in mode space. This reduction will not be used as the primary numerical solver, but it is useful for checking the structure of the biased correlated random walk and for verifying that the unbiased CRW operator is recovered when the bias is removed. Let $\delta(r,\theta)$ denote the bias phase in the turning kernel, and expand
\[
T(r,\theta,\phi)
=
\sum_{m,n\in\mathbb Z}
U_{m,n}(r)\,\mathrm{e}^{im\theta}\mathrm{e}^{in\phi}.
\]
For fixed heading mode $n$, the factor $\mathrm{e}^{in\delta(r,\theta)}$ introduces spatial dependence through the polar angle $\theta$. If $C_{n,q}(r)$ denotes its Fourier coefficients in $\theta$, then the turning part of the BCRW operator becomes
\begin{equation}
(\mathcal L_{\mathrm{turn}} U)_{m,n}(r)
=
\mu\left[
\sum_{q\in\mathbb Z}
\frac{I_n(\kappa)}{I_0(\kappa)}
C_{n,q}(r)\,
U_{m-q,n}(r)
-
U_{m,n}(r)
\right].
\label{eq:bcrw-fourier-turn}
\end{equation}
Thus the biased turning operator preserves the heading mode $n$, but mixes the position modes $m$ through a discrete convolution. This is the Fourier-mode-space expression of the fact that the external cue introduces spatially dependent directional preference, in contrast to the unbiased CRW where the turning kernel depends only on the turning angle \cite{Alt1980,CodlingPlankBenhamou2008,OthmerHillen2002}.

In the unbiased limit $\delta\equiv 0$, the coefficients satisfy $C_{n,0}=1$ and $C_{n,q}=0$ for $q\neq 0$. The convolution then collapses, and \eqref{eq:bcrw-fourier-turn} reduces to the diagonal CRW operator in \eqref{eq:K-fourier-crw}. The absorbing condition gives $U_{m,n}(a)=0$ for all $m,n$, while specular reflection at the outer boundary gives
\[
U_{m,n}(R)=(-1)^n U_{m+2n,-n}(R).
\]
The full coupled BCRW radial system, including both streaming and turning contributions, is shown in \Cref{app:bcrw}.

\paragraph{Boundary-layer correction to the diffusion limit MFPT}
\label{sec:analysis-eps}
We can use the moment-closure framework for velocity-jump processes
\cite{HillenOthmer2000,OthmerHillen2002} to identify the first correction
to the diffusion-limit MFPT. Let
\[
\lambda=\mu(1-m_1),
\qquad
m_1=\frac{I_1(\kappa)}{I_0(\kappa)},
\]
where $\lambda$ is the relaxation rate of the first directional moment. We
expand the CRW mean first-passage time in powers of $1/\lambda$,
\[
T_{\mathrm{CRW}}(r;\kappa)
=
T_0(r) + \frac{1}{\lambda}\,T_1(r) + \mathrm{H.O.T.},
\]
where $T_0$ is the leading-order (diffusion-limit) MFPT and $T_1$ is the
first correction. At leading order the CRW reduces to a diffusion process
with effective diffusivity
\[
D_{\mathrm{eff}}
=
\frac{s^2}{2\lambda}
=
\frac{s^2}{2\mu(1-m_1)},
\]
and $T_0$ satisfies
\[
D_{\mathrm{eff}}\,\Delta T_0 = -1,
\qquad
T_0(a)=0,
\qquad
\partial_r T_0(R)=0,
\]
so that $T_0 = T_{\mathrm{SRW}}^{D_{\mathrm{eff}}}(r)$ from
\eqref{eq:srw}. The next-order term $T_1$ satisfies
\[
D_{\mathrm{eff}}\,\Delta T_1 = D_{\mathrm{eff}}^2\,\Delta^2 T_0.
\]
Because $\Delta^2 T_0=0$ in the interior, we can see that
$\Delta T_1=0$ inside the annulus. Specular reflection at $r=R$ is an exact
transport boundary condition that is inherited order by order, giving
$\partial_r T_1(R)=0$. Together these facts imply that $T_1$ is constant in $r$ and its
value is not set by bulk forcing but by the kinetic boundary layer at the
absorbing target, which converts the half-range absorbing condition
$T(a,\psi)=0$ for inward headings into an effective scalar condition on the
outer diffusion solution. The full calculation is given in
\Cref{app:eps-correction} and yields
\begin{equation}
T_{\mathrm{CRW}}(r;\kappa)
=
T_{\mathrm{SRW}}^{D_{\mathrm{eff}}}(r)
+
\mathcal{F}_{\mathrm{abs}}(a,R,s;\kappa)
+
\mathrm{H.O.T.},
\label{eq:milne-correction}
\end{equation}
where
\begin{equation}
\mathcal{F}_{\mathrm{abs}}(a,R,s;\kappa)
=
\chi_{\mathrm{abs}}(\kappa)\,\frac{R^2-a^2}{as}.
\label{eq:Fabs}
\end{equation}
Here $\chi_{\mathrm{abs}}(\kappa)$ is a dimensionless extrapolation constant
determined by the half-range absorbing boundary condition and the turning
kernel. This form predicts that the leading finite-persistence correction
shifts the radial MFPT profile by a $\kappa$-dependent constant rather than
changing its radial shape, in analogy with the Milne extrapolation length that
appears in mean-exit problems when a kinetic boundary layer replaces a sharp
absorbing diffusion boundary condition~\cite{HaganDoeringLevermore1989}.
%%%%%%%%%%%%%%%%%%%%%%%%%%%%%%%%%%%%%%%%%%%%%%%%%%
\section{Numerical methods}\label{sec:numerics}
\begin{figure}[htbp]
  \centering
  \includegraphics[width=0.5\textwidth]{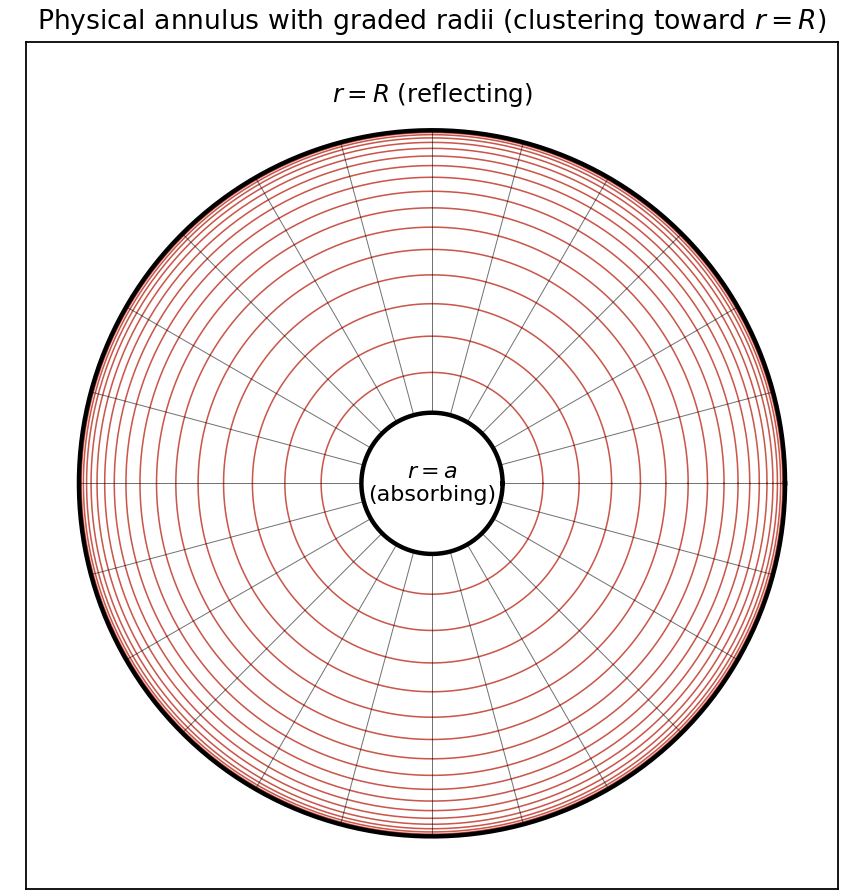}
  \caption{Power-law graded radial mesh on the annulus, shown at
    $\beta = 3$, $N_r = 41$, $N_\psi = 36$ for visual clarity (the
    production runs use larger $N_r, N_\psi$; see
    \Cref{sec:mesh-calibration}). Radial cells cluster toward the
    reflecting outer boundary $r = R$ in order to resolve the
    persistence-induced wall layer of width
    $w(\kappa) \approx 5.41\,\kappa^{-1.27}$, while the angular grid
    remains uniform. The formula givesa uniform radial mesh when
    $\beta = 1$.}
  \label{fig:graded-annulus}
\end{figure}

\paragraph{Domain discretisation and graded
mesh}\label{sec:domain-disc}
We discretise the annulus using a radial grid
$a=r_1<r_2<\dots<r_{N_r}=R$ and a uniform angular grid
$\psi_j=2\pi(j-1)/N_\psi$, $j=1,\dots,N_\psi$. The radial nodes are placed according to a power-law, clustering toward the reflecting boundary (\Cref{fig:graded-annulus}),
\begin{equation}
r_i=a+(R-a)\bigl(1-(1-\xi_i)^\beta\bigr),\qquad
\xi_i=(i-1)/(N_r-1).
\label{eq:graded-mesh}
\end{equation}

The clustering exponent $\beta$ is set empirically to resolve the
persistence-induced kinetic boundary layers, which narrow with $\kappa$
at both the absorbing inner boundary and the reflecting outer
boundary. \Cref{sec:mesh-calibration} reports the sweep that selects
$\beta$ and explains why a $\kappa$-dependent choice is used in the
production runs.

\paragraph{Semi-Lagrangian $(r,\psi)$ solver}\label{sec:numerics-sl}
The streaming part of \Cref{eq:crw-backward} is handled by tracing the
characteristic ODEs $\mathrm{d} r/\mathrm{d} t = s\cos\psi$,
$\mathrm{d}\psi/\mathrm{d} t = -s\sin\psi/r$ backward over a pseudo-time step
$\Delta\tau$ to a foot-point $(r_d,\psi_d)$. If the trajectory crosses
$r=a$ the absorbing condition $T=0$ is enforced, and if it crosses
$r=R$ the velocity is reflected $\psi\mapsto\pi-\psi$, followed by an inward nudge of $10^{-9}\,(R-a)$ to keep the foot point
strictly inside the domain. This is followed by exact integration of the remaining time step.
The interpolated value of the previous iterate at the foot-point
defines $S[T^{(m)}]_{i,j}$. The turning integral is approximated by
the trapezoidal rule on the uniform angular grid
\cite{Trefethen2014}, and an implicit treatment of turning combined
with semi-Lagrangian streaming yields the fixed-point update
\begin{equation}
\Bigl(\tfrac{1}{\Delta\tau}+\mu\Bigr)T_{i,j}^{(m+1)}
-\mu\!\sum_{\ell=1}^{N_\psi}w_\ell K^*(\psi_j,\psi_\ell)\,T_{i,\ell}^{(m+1)}
=\tfrac{1}{\Delta\tau}\,S[T^{(m)}]_{i,j}-1.
\label{eq:sl-update}
\end{equation}
At each radial level this is an $N_\psi\times N_\psi$ linear system in
the angular unknowns. We iterate until the discrete residual norm $\|\mathbf{r}\|_\infty$ falls below
$10^{-8}$ in MFPT units. The angle-averaged MFPT after
convergence is computed as
$\overline{T}(r_i)=(2\pi)^{-1}\sum_\ell w_\ell T_{i,\ell}$.
\begin{figure}[t]
\centering
\begin{tikzpicture}[
    >={Stealth[length=2.2mm]},
    grid pt/.style       = {circle, fill=black, inner sep=1.1pt},
    arrival/.style       = {circle, draw=blue!70!black, fill=blue!20, thick, inner sep=2pt},
    foot/.style          = {circle, draw=red!80!black,  fill=red!20,  thick, inner sep=2pt},
    hit/.style           = {circle, draw=orange!85!black, fill=orange!30, thick, inner sep=1.7pt},
    cell/.style          = {fill=yellow!25, draw=orange!75, dashed, thick},
    trace/.style         = {thick, red!80!black,
        decoration={markings, mark=at position 0.55 with {\arrow{>}}},
        postaction={decorate}},
    unconstrained/.style = {thin, gray!60, dashed},
]

% ============================================================
% Panel (a): interior backward step in (r, psi)
% ============================================================
\begin{scope}
    % axes
    \draw[->, thick] (0.5,0.5) -- (6.2,0.5) node[below right] {$r$};
    \draw[->, thick] (0.5,0.5) -- (0.5,5.0) node[above left] {$\psi$};

    % grid nodes
    \foreach \i in {1,...,5}{%
        \foreach \j in {1,...,4}{%
            \node[grid pt] at (\i,\j) {};%
        }
    }

    % interpolation cell
    \draw[cell] (3,2) rectangle (4,3);
    \node[font=\scriptsize, orange!80!black] at (3.5,1.72) {interpolation cell};

    % arrival and foot points
    \node[arrival] (A) at (4,3) {};
    \node[foot] (F) at (3.38,2.46) {};

    % clean labels away from the points
    \node[blue!70!black, font=\small, anchor=west] at (4.18,3.18) {$(r_i,\psi_j)$};
    \node[red!80!black, font=\small, anchor=east] at (3.20,2.32) {$(r_d,\psi_d)$};

    % backward characteristic
    \draw[trace] (A) to[bend right=14] (F);

    % small delta tau label, moved away from the arrowhead
    \node[red!80!black, font=\scriptsize] at (3.93,2.63) {$\Delta\tau$};

    % four corners emphasized lightly
    \foreach \p in {(3,2),(4,2),(3,3),(4,3)}{
        \node[hit, inner sep=0.9pt] at \p {};
    }

    % subcaption
    \node[align=center, font=\scriptsize] at (3.45,-0.55) {(a) Interior backward step};
\end{scope}

% ============================================================
% Panel (b): reflection at the outer wall
% ============================================================
\begin{scope}[xshift=8.4cm, yshift=-0.5cm]
    % outer wall
    \draw[thick] (25:4.5) arc (25:88:4.5);
    \node[font=\scriptsize] at (29:4.9) {$r=R$};

    % points
    \coordinate (Ain)  at (61:3.75);
    \coordinate (X)    at (61:4.5);
    \coordinate (Xout) at (61:5.25);
    \coordinate (Foot) at (52:3.35);

    % % unconstrained path, kept light and separated
    % \draw[unconstrained] (X) -- (Xout);
    % \node[font=\scriptsize, gray!70!black, anchor=west] at ($(Xout)+(0.18,0.12)$) {unconstrained};

    % incoming and reflected characteristic pieces
    \draw[trace] (Ain) -- (X);
    \draw[trace] (X) -- (Foot);

    % points
    \node[arrival] at (Ain) {};
    \node[hit] at (X) {};
    \node[foot] at (Foot) {};

    % labels placed away from geometry
    \node[blue!70!black, font=\scriptsize, anchor=east] at ($(Ain)+(-0.20,0.10)$) {$(r_i,\psi_j)$};
    \node[orange!85!black, font=\scriptsize, anchor=east] at ($(X)+(-0.3,0)$) {hit at $t^*$};
    \node[red!80!black, font=\scriptsize, anchor=east] at ($(Foot)+(-0.18,-0.05)$) {$(r_d,\psi_d)$};

    % outward normal
    \draw[->, black, thin] (X) -- ($(X)+(61:0.65)$);
    \node[black, font=\scriptsize] at ($(X)+(61:0.88)$) {$\hat{\mathbf n}$};

    % remaining time label moved below the reflected segment
    \node[font=\scriptsize, red!80!black, align=center] at ($(Foot)+(0.55,-0.40)$)
    {remaining\\[-1mm]$\Delta\tau-t^*$};

    % reflection formula in a clean location to the right
    \node[font=\scriptsize, align=left, anchor=west] at (3.45,3.95)
    {$\psi\mapsto \pi-\psi$\\
     nudge inward by $10^{-9}$};

    % subcaption
    \node[align=center, font=\scriptsize] at (1.9,-0.02) {(b) Outer-wall reflection};
\end{scope}

\end{tikzpicture}

\caption{Semi-Lagrangian backward-characteristic step.
(a) From a grid node $(r_i,\psi_j)$, the characteristic equation is integrated backward by $\Delta\tau$ to an off-grid foot point $(r_d,\psi_d)$. The value at the foot point is obtained by bilinear interpolation from the surrounding cell.
(b) If the backward characteristic reaches the outer wall $r=R$ before the end of the step, the velocity is specularly reflected according to
$\mathbf v\mapsto \mathbf v-2(\mathbf v\cdot\hat{\mathbf n})\hat{\mathbf n}$,
equivalently $\psi\mapsto \pi-\psi$, and the remaining time $\Delta\tau-t^*$ is integrated to the final foot point.}
\label{fig:semi-lagrangian-step}
\end{figure}
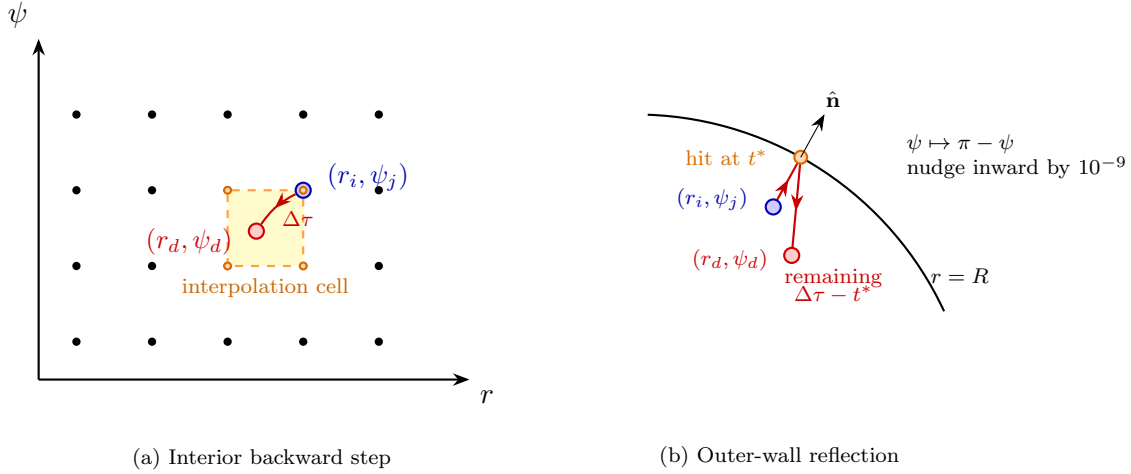

\paragraph{Time step-free stationary
formulation}\label{sec:numerics-stationary}
The semi-Lagrangian iteration carries an $O(\mu\Delta\tau)$ accuracy
ceiling, characterized in \Cref{sec:mesh-calibration}. To remove
it, we solve \Cref{eq:crw-backward} directly as a single sparse linear
system. With $u_{i,j}\approx T(r_i,\psi_j)$ and an upwind treatment
of the radial advection (sign of $\cos\psi_j$),
\begin{equation}
s\cos\psi_j\,(\partial_r T)_{i,j}
\;\approx\; s\cos\psi_j\cdot
\begin{cases}
(u_{i,j}-u_{i-1,j})/h_i^- & \cos\psi_j>0,\\
(u_{i+1,j}-u_{i,j})/h_i^+ & \cos\psi_j<0,
\end{cases}
\label{eq:upwind}
\end{equation}
combined with a centred difference for the $\partial_\psi T$ term and
a boundary-aware split-Simpson rule for the angular quadrature
\cite{QuarteroniSaccoSaleri2007}, yields a sparse
$(N_r N_\psi)\times(N_r N_\psi)$ block-banded linear system. Specular
reflection at $r=R$ is enforced as a permutation block coupling
$\psi_j$ to $\pi-\psi_j$, and the absorbing condition at $r=a$ is
imposed by row replacement. The system is solved in one shot by a
sparse direct solver. 

\paragraph{BCRW solver extension}\label{sec:numerics-bcrw}
The BCRW kernel \eqref{eq:bcrw-kernel} depends on $(r,\theta)$
through the bias phase $\phi_0(r,\theta)$, so the angular system at
each spatial node is no longer translation-invariant in $\psi$. We
keep the semi-Lagrangian streaming step from \Cref{sec:numerics-sl}
applied to the full $(r,\theta,\phi)$ field and replace the angular
turning solve by an $N_\phi\times N_\phi$ system whose matrix entries
are the position-dependent samples of $K^*$. 

As a simple representative
chemotactic field we will consider radial bias toward the absorbing target,
$\phi_0(r,\theta)=\theta+\pi$ (inward radial direction), so that the
problem retains rotational symmetry in $\theta$ and the unknown
reduces to $T(r,\psi)$ with a $\psi$-only kernel that nonetheless
breaks $\psi\to-\psi$ symmetry.

The position-dependent quadrature was verified against the unbiased
CRW solver in the $d_\tau \to 0$ limit. At
$d_\tau = 0$, the BCRW field reproduces the unbiased CRW field to
roundoff error, confirming
that the position-dependent angular system collapses to the
translation-invariant CRW system in the unbiased limit. For
$d_\tau > 0$ at the production grid ($\beta = 3$, $N_r = 121$,
$N_\psi = 144$), the maximum probe relative deviation
$|\bar T_{\mathrm{BCRW}} - \bar T_{\mathrm{CRW}}| / \bar T_{\mathrm{CRW}}$
scales linearly in $d_\tau$ with fitted slopes $0.997$ ($\kappa = 1$)
and $0.990$ ($\kappa = 3$), consistent with an analytical leading
order expansion of the kernel. The deviation crosses the
semi-Lagrangian structural ceiling of $2.7\%$ near $d_\tau \approx
10^{-2}$ at both tested $\kappa$. We use this to set the bias range over which
the BCRW solver can be compared directly with the unbiased CRW
reference without being dominated by solver error. The full $d_\tau$ sweep is shown in the supplementary material (\Cref{fig:SI-beta-sweep}).
The unbiased limit
$d_\tau=0$ recovers \Cref{sec:numerics-sl}. We use the small-$d_\tau$
sweep as a consistency check before reporting any chemotactic
prediction.

\paragraph{Event-driven Monte Carlo simulation}\label{sec:numerics-mc}
Between turning events the walker advances ballistically along
$\mathrm{d} \mathbf{x}/\mathrm{d}t=s(\cos\theta,\sin\theta)$, with turning events
generated as a Poisson process with rate $\mu$ and new headings drawn
from \Cref{eq:vonmises} via the Best--Fisher rejection method
\cite{Best1979}. Specular reflection is applied at $r=R$. From $M$
independent realisations,
$\widehat{T}(r_0,\psi_0) = M^{-1}\sum_{m=1}^M \tau^{(m)}$ with the
sample standard error reported as a $95\%$ confidence interval. The
event-driven simulator carries no $\Delta t$ discretisation error and
serves as the primary cross-validator of the deterministic solvers.
%%%%%%%%%%%%%%%%%%%%%%%%%%%%%
\section{Verification and convergence}\label{sec:verification}

We verified the numerical workflow by comparing the direct \((r,\psi)\) solver with the independent event-driven Monte Carlo simulator. Before this comparison, we checked the main building blocks of the solver separately: the discrete von Mises kernel preserves its normalization and first trigonometric moment, the absorbing boundary at \(r=a\) is imposed correctly, and the specular reflection rule at \(r=R\) reproduces the analytical reflected trajectory. In the semi-Lagrangian solver, the fixed-point iteration was continued
until the absolute infinity-norm update satisfied
$\|T^{(m+1)} - T^{(m)}\|_\infty \leq 10^{-8}$ in MFPT units. 
The main validation is based on the relative gap
\[
g = \frac{\left|\bar T_{\mathrm{direct}} - \widehat T_{\mathrm{MC}}\right|}
         {\widehat T_{\mathrm{MC}}},
\]
reported throughout as a percentage difference $100\,g\,(\%)$, where
$\bar T_{\mathrm{direct}}$ is the angle-averaged direct-solver MFPT
and $\widehat T_{\mathrm{MC}}$ is the Monte Carlo estimate. The comparison was performed at four representative radii,
\[
r\in\{1.5,2.5,3.5,5\},
\]
and for \(\kappa\in\{0,0.5,1,2,3,4,5\}\). 

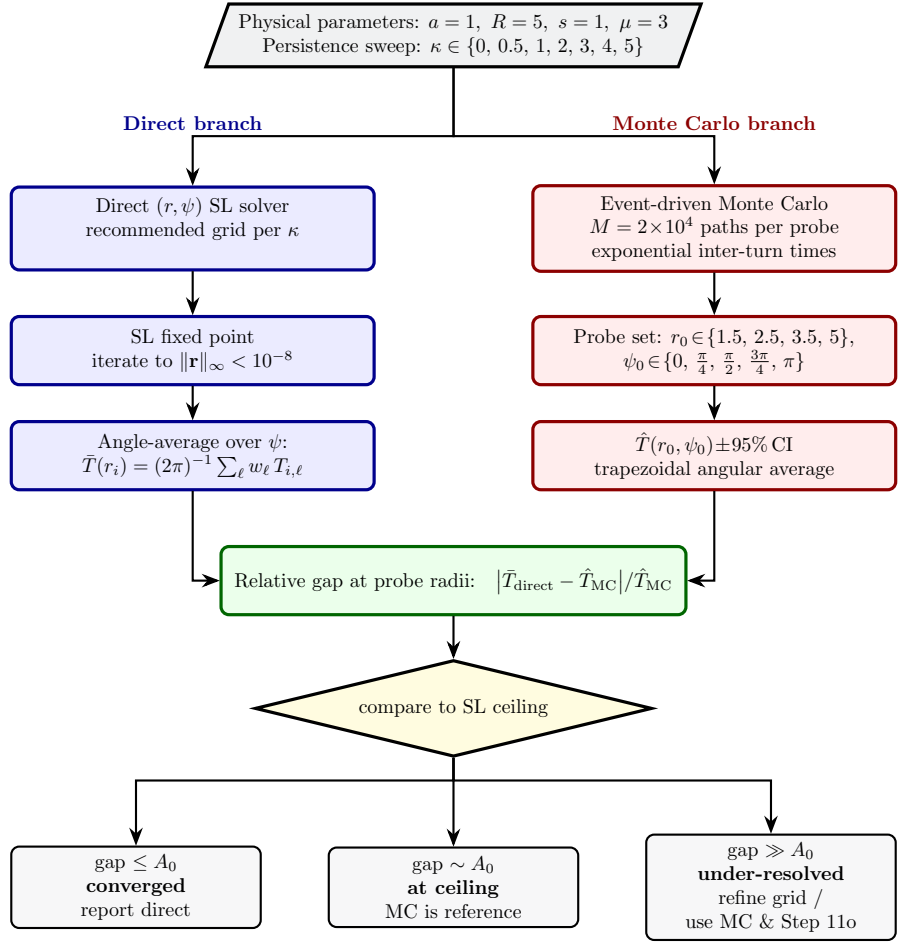
\begin{figure}[t]
	\centering
	\begin{tikzpicture}[
    scale=0.75,
    transform shape,
		> = {Stealth[length=2.4mm]},
		font   = \small,
		io/.style       = {trapezium, trapezium left angle=70, trapezium right angle=110,
			draw=black, very thick, align=center, fill=gray!12,
			inner sep=5pt, minimum height=10mm, minimum width=80mm},
		direct/.style   = {rectangle, rounded corners=3pt, draw=blue!55!black, very thick,
			align=center, inner sep=5pt, minimum height=12mm,
			minimum width=64mm, fill=blue!8},
		mc/.style       = {rectangle, rounded corners=3pt, draw=red!55!black,  very thick,
			align=center, inner sep=5pt, minimum height=12mm,
			minimum width=64mm, fill=red!7},
		cmp/.style      = {rectangle, rounded corners=3pt, draw=green!40!black, very thick,
			align=center, inner sep=5pt, minimum height=12mm,
			minimum width=82mm, fill=green!8},
		decision/.style = {diamond, draw=black, very thick, aspect=2.6, align=center,
			inner sep=1pt, fill=yellow!15, minimum width=70mm,
			minimum height=14mm},
		outcome/.style  = {rectangle, rounded corners=3pt, draw=black, thick,
			align=center, inner sep=4pt, minimum height=14mm,
			minimum width=44mm, fill=gray!6},
		arrow/.style    = {-{Stealth[length=2.4mm]}, thick},
		]
		
		% ============== Inputs (top) ==============
		\node[io] (inp) at (0, 0)
		{Physical parameters: $a=1,\ R=5,\ s=1,\ \mu=3$\\
			Persistence sweep: $\kappa\in\{0,\,0.5,\,1,\,2,\,3,\,4,\,5\}$};
		
		% ============== Branch heads ==============
		\node[direct] (d1) at (-4.6, -3.4)
		{Direct $(r,\psi)$ SL solver\\
			recommended grid per $\kappa$\\
			};
		
		\node[mc] (m1) at ( 4.6, -3.4)
		{Event-driven Monte Carlo\\
			$M=2\!\times\!10^{4}$ paths per probe\\
			exponential inter-turn times};
		
		% ============== Direct branch ==============
		\node[direct] (d2) at (-4.6, -5.55)
		{SL fixed point \\
			iterate to $\|\mathbf r\|_\infty<10^{-8}$};
		
		\node[direct] (d3) at (-4.6, -7.4)
		{Angle-average over $\psi$:\\
			$\bar T(r_i)=(2\pi)^{-1}\sum_\ell w_\ell\,T_{i,\ell}$};
		
		% ============== MC branch ==============
		\node[mc] (m2) at ( 4.6, -5.55)
		{Probe set: $r_0\!\in\!\{1.5,\,2.5,\,3.5,\,5\}$,\\
			$\psi_0\!\in\!\{0,\,\tfrac{\pi}{4},\,\tfrac{\pi}{2},\,\tfrac{3\pi}{4},\,\pi\}$};
		
		\node[mc] (m3) at ( 4.6, -7.4)
		{$\hat T(r_0,\psi_0)\!\pm\!95\%\,\mathrm{CI}$\\
			trapezoidal angular average};
		
		% ============== Comparison ==============
		\node[cmp] (cmp) at (0, -9.6)
		{Relative gap at probe radii:\quad
			$\bigl|\bar T_{\mathrm{direct}}-\hat T_{\mathrm{MC}}\bigr|/\hat T_{\mathrm{MC}}$};
		
		% ============== Decision ==============
		\node[decision] (dec) at (0, -11.85)
		{compare to SL ceiling};
		
		% ============== Outcomes ==============
		\node[outcome] (oA) at (-5.6, -15)
		{gap $\le A_0$\\\textbf{converged}\\report direct};
		\node[outcome] (oB) at ( 0.0, -15)
		{gap $\sim A_0$\\\textbf{at ceiling}\\MC is reference};
		\node[outcome] (oC) at ( 5.6, -15)
		{gap $\gg A_0$\\\textbf{under-resolved}\\refine grid /\\use MC \& Step~11o};
		
		% ============== Edges ==============
		\draw[arrow] (inp.south) -- ++(0,-1.2) -| (d1.north);
		\draw[arrow] (inp.south) -- ++(0,-1.2) -| (m1.north);
		\draw[arrow] (d1) -- (d2);
		\draw[arrow] (d2) -- (d3);
		\draw[arrow] (m1) -- (m2);
		\draw[arrow] (m2) -- (m3);
		\draw[arrow] (d3.south) |- (cmp.west);
		\draw[arrow] (m3.south) |- (cmp.east);
		\draw[arrow] (cmp) -- (dec);
		\draw[arrow] (dec.south) -- ++(0,-0.4) -| (oA.north);
		\draw[arrow] (dec.south) -- ++(0,-0.4) -| (oB.north);
		\draw[arrow] (dec.south) -- ++(0,-0.4) -| (oC.north);
		
		% ============== Branch labels ==============
		\node[blue!55!black, font=\bfseries\small] at (-4.6, -1.55) {Direct branch};
		\node[red!55!black,  font=\bfseries\small] at ( 4.6, -1.55) {Monte Carlo branch};
		
	\end{tikzpicture}
	\caption{Validation workflow. The direct \((r,\psi)\) solver and the event-driven Monte Carlo simulator are run independently at the same parameter values. Their angle-averaged MFPTs are compared using the relative gap. Small gaps indicate convergence, gaps near the observed semi-Lagrangian ceiling indicate the limitation of the pseudo-time formulation, and larger gaps indicate under resolution or transition to the strong persistence regime.}
\label{fig:validation_flowchart}
\end{figure}

\begin{figure}[t]
\centering

\begin{subfigure}{0.52\textwidth}
    \centering
    \includegraphics[width=\textwidth]{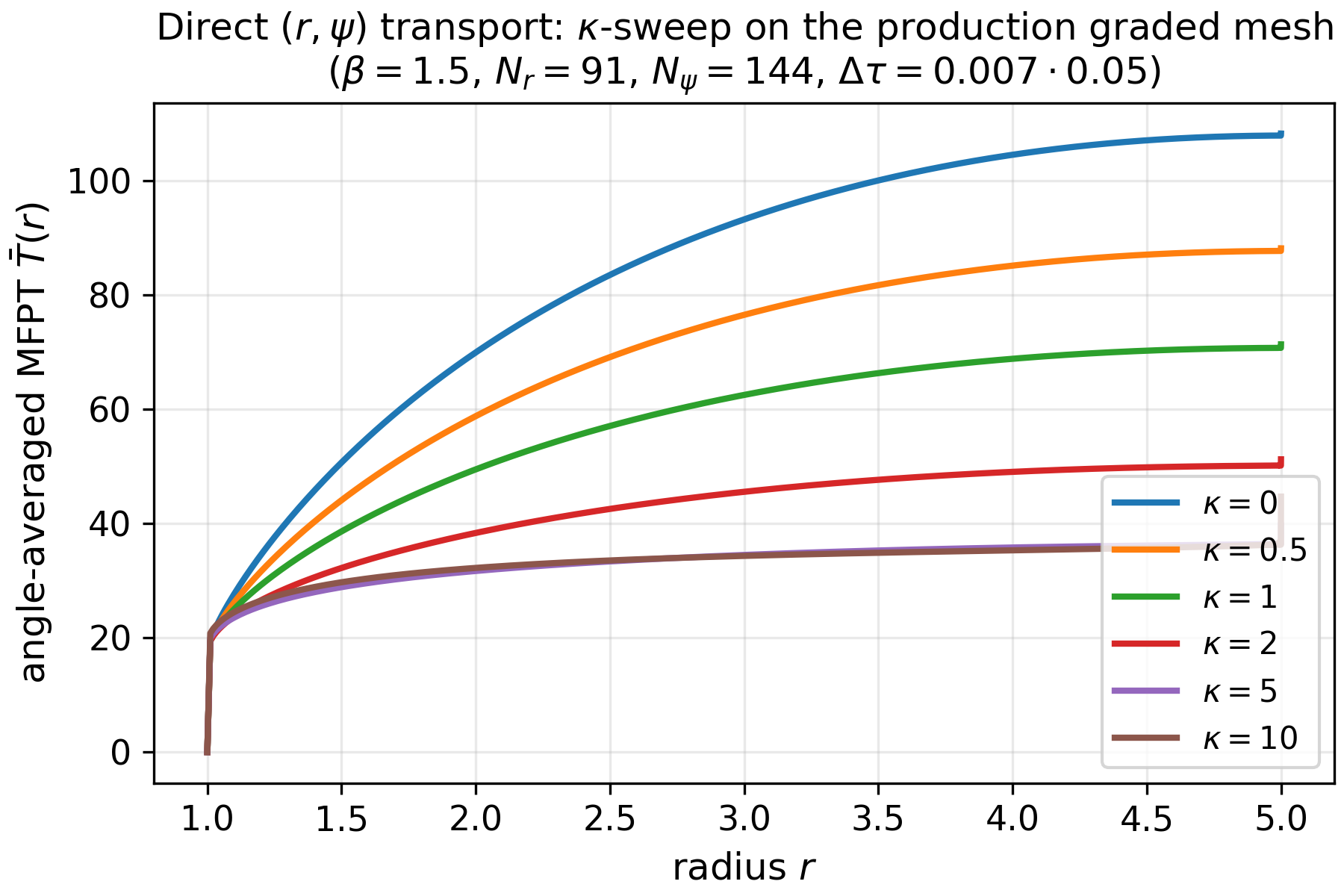}
    \caption*{\textbf{(a)}}
\end{subfigure}
\hfill
\begin{subfigure}{0.46\textwidth}
    \centering
    \includegraphics[width=\textwidth]{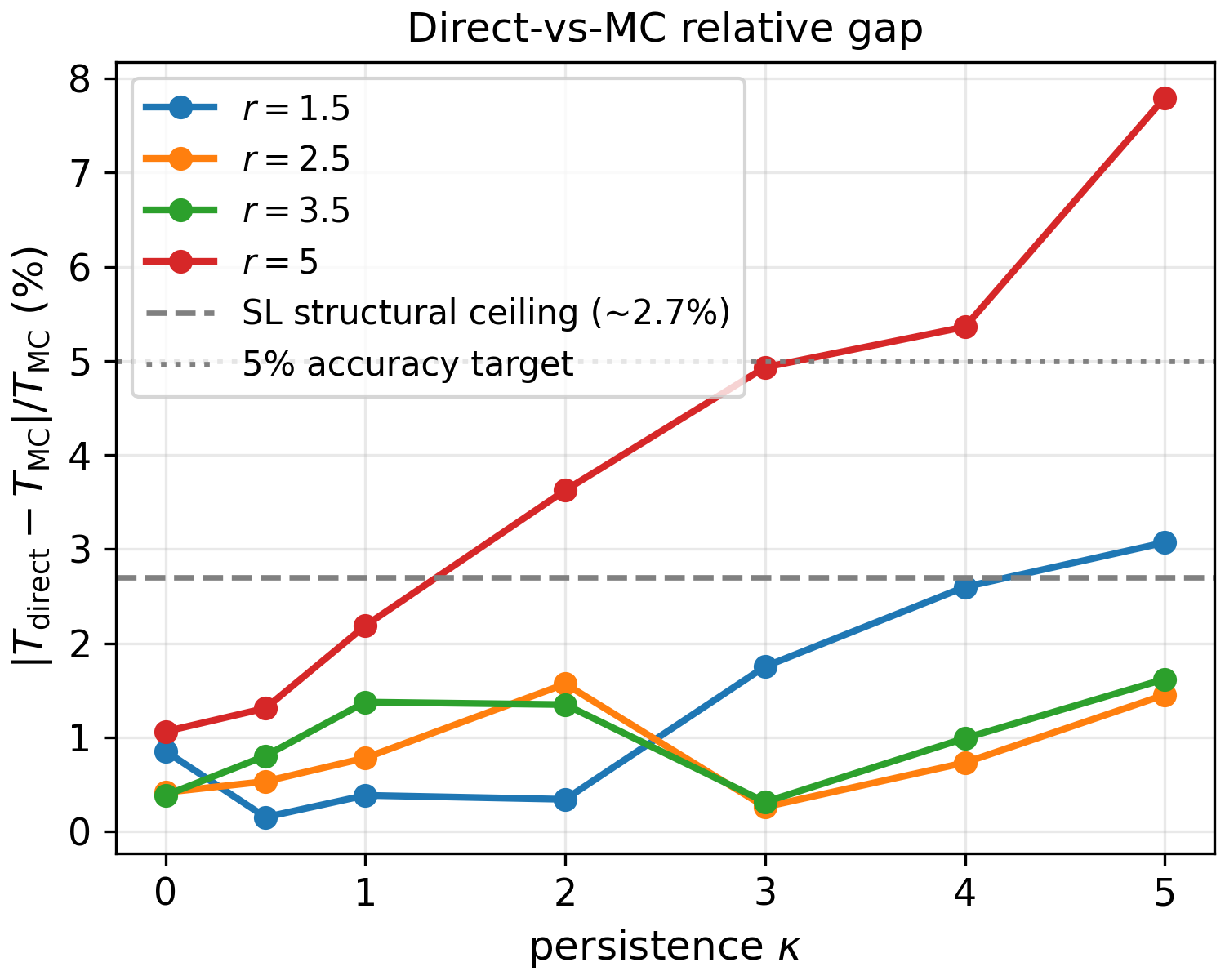}
    \caption*{\textbf{(b)}}
\end{subfigure}

\caption{Persistence sweep and validation of the direct semi-Lagrangian
  solver in the annulus $a < r < R$ with absorbing inner boundary at
  $r = a$ and specularly reflecting outer boundary at $r = R$.
  Parameters: $a = 1$, $R = 5$, walker speed $s = 1$, turning rate
  $\mu = 3$. Grid: graded mesh with clustering exponent
  $\beta$ and radial/angular node counts $(N_r, N_\psi)$ chosen per
  $\kappa$ following~\eqref{eq:graded-mesh}; pseudo-time step
  $\Delta\tau \in [0.007, 0.05]$.
  \textbf{(a)} Angle-averaged MFPT profiles from the
  direct $(r, \psi)$ transport solver across the persistence sweep
  $\kappa \in \{0, 0.5, 1, 2, 3, 4, 5\}$. Increasing $\kappa$
  decreases the MFPT over the validated range $\kappa \in [0, 5]$.
  The $\kappa = 10$ curve is shown only to illustrate the onset of
  the high-persistence regime, where the semi-Lagrangian solver is
  no longer used as the primary reference.
  \textbf{(b)} Direct-vs-Monte-Carlo percentage difference
  $100\,|\bar T_{\mathrm{direct}} - \widehat T_{\mathrm{MC}}| /
  \widehat T_{\mathrm{MC}}$ at the four probe radii
  $r_0 \in \{1.5, 2.5, 3.5, 5\}$, evaluated against the event-driven
  Monte Carlo reference ($M = 2 \times 10^4$ paths per probe;
  see \Cref{sec:numerics}). The dashed line marks the representative Richardson-extrapolated
ceiling $A_0 \approx 2.7\%$ at $(\kappa, r) = (5, 5)$, shown for
reference. The largest gaps occur near
the reflecting wall at $r = 5$ for $\kappa \ge 2$ and additionally
at the absorbing inner probe $r = 1.5$ for $\kappa \ge 3$, reflecting
the kinetic boundary layers at the two boundaries.}
\label{fig:kappa-sweep-validation}
\end{figure}

The validation workflow is shown in \Cref{fig:validation_flowchart}, and the resulting gaps are reported in \Cref{tab:gap-grid} and shown in \Cref{fig:kappa-sweep-validation}.

Our extensive validation $(M = 10^6$ MC paths per run) reveals
that the semi-Lagrangian solver has a structural accuracy floor that
varies with both persistence and radius $r$ rather than a single
global value. The gap is small at interior radii for $\kappa \le 1$
($\lesssim 1.5\%$) and grows toward the outer wall at every $\kappa$,
reaching $5.0\%$ at $(\kappa, r) = (2, 5)$ and $6.5\%$ at
$(\kappa, r) = (5, 5)$ on the production grid
\eqref{eq:graded-mesh}. 
\subsection{Outer-wall layer and calibration of the graded mesh}
\label{sec:mesh-calibration}

The semi-Lagrangian solver has two kinetic boundary layers whose
width narrows with $\kappa$: one at the absorbing inner boundary
$r = a$ and one at the
reflecting outer boundary $r = R$ (persistence-induced wall layer).
A fit to the layer width and amplitude on uniform grids gives
\begin{equation}
  w(\kappa) \approx 5.41\, \kappa^{-1.27},
  \qquad
  A(\kappa) \approx 2.95\, \kappa^{\,0.835},
  \label{eq:wall-layer-fit}
\end{equation}
over $\kappa \in \{0.5, 1, 2, 5, 10\}$. The layer narrows and
strengthens with persistence, so the chosen mesh must be both
$\kappa$-dependent and graded.

The adaptive mesh reduces to a uniform radial mesh at
$\beta = 1$ and concentrates resolution at $r = R$ as $\beta$ grows.
A sweep over $\beta \in \{1.5, 2, 3, 5\}$ and
$N_r \in \{91, 121, 181, 241, 361, 541\}$ at $N_\psi = 144$,
$\Delta\tau = 0.007$ identified the candidate
production grid~\eqref{eq:graded-mesh}. The low-$\kappa$ branch
uses a gentle grading ($\beta = 1.5$) with a refined radial grid
($N_r = 361$) to resolve the inner Milne layer that dominates at low
persistence, while the high-$\kappa$ branch uses strong outer clustering
($\beta = 3$) at a moderate radial count ($N_r = 241$) to resolve the
narrowing outer wall layer. Comparison against extensive Monte
Carlo simulation ($M = 10^6$ runs) on this grid is reported in \Cref{tab:gap-grid}.
The residual SL-vs-MC gap reflects the $O(\mu\Delta\tau)$ structural
ceiling of the pseudo-time formulation rather than unresolved layer
structure, and is removed in the stationary solver.

\begin{table}[htbp]
  \centering
  \caption{Direct-vs-Monte-Carlo percentage difference
    $100\,|\bar T_{\mathrm{direct}} - \widehat T_{\mathrm{MC}}| /
    \widehat T_{\mathrm{MC}}$. The gap defines the
    $(\kappa, r)$-dependent semi-Lagrangian structural floor.
    Largest values appear at the reflecting outer boundary ($r = 5$)
    for $\kappa \ge 2$ and at the absorbing inner probe ($r = 1.5$)
    for $\kappa \ge 3$, reflecting the two kinetic boundary layers.
    Values are given to two decimal places.}
  \label{tab:gap-grid}
  \begin{tabular}{lcccc}
    \toprule
    & \multicolumn{4}{c}{probe radius $r$} \\
    \cmidrule(lr){2-5}
    $\kappa$ & $1.5$ & $2.5$ & $3.5$ & $5$ \\
    \midrule
    $0$   & $0.82$ & $0.39$ & $0.08$ & $0.35$ \\
    $0.5$ & $0.75$ & $0.23$ & $0.69$ & $1.38$ \\
    $1$   & $0.29$ & $1.35$ & $1.41$ & $2.65$ \\
    $2$   & $1.12$ & $2.13$ & $2.49$ & $5.01$ \\
    $3$   & $4.05$ & $2.23$ & $2.13$ & $2.21$ \\
    $4$   & $4.16$ & $2.43$ & $2.17$ & $4.42$ \\
    $5$   & $4.17$ & $2.55$ & $2.33$ & $6.55$ \\
    \bottomrule
  \end{tabular}
\end{table}

%%%%%%%%%%%%%%%%%%%%%%%%%%%%%%%%%%%%%
\section{Biological calibration and predictions}\label{sec:bio}
\subsection{Mapping the concentration parameter to cell migration data}\label{sec:bio-mapping}

To connect the numerical persistence sweep to biological motility data, we use the one-step directional persistence
\[
c=\langle \cos\Delta\theta\rangle = \frac{I_1(\kappa)}{I_0(\kappa)}.
\]
This mapping is monotone: \(c=0\) corresponds to isotropic reorientation, while \(c\to 1\) corresponds to highly persistent motion. Representative cell-tracking studies place many biological walkers in the range
\[
c\approx 0.3\text{--}0.9,
\qquad
\kappa\approx 0.7\text{--}5.
\]
This includes run-and-tumble \emph{E.~coli}, lymphocytes and NK-cell-like motion in lymph nodes, neutrophils, monocytes, fibroblasts, and other mesenchymal cells \cite{BajenoffEtAl2006,BeltmanEtAl2007,Berg2004,BergBrown1972,GailBoone1970,JannatEtAl2010,MaiuriEtAl2015,MillerEtAl2002,MillerEtAl2003}. More strongly persistent cells, such as \emph{Dictyostelium} and keratocytes, can extend beyond this range and motivate the strong persistence discussion in \Cref{sec:analysis-strong} \cite{EuteneuerSchliwa1984,KerenEtAl2008,LeeEtAl1993,VanHaastertBosgraaf2009} and in the ecological random walk literature \cite{CodlingThesis2003,CodlingPlankBenhamou2008}. We therefore take
\[
\kappa\in(0,5]
\]
as the main biologically relevant and numerically validated range for the present study.

\begin{table}[t]
\caption{Approximate mapping between representative biological motility classes and the von Mises persistence parameter. Values are approximate because the underlying studies use different sampling intervals, imaging environments, and motility summaries.}
\label{tab:cell-types}
\centering\small
\begin{tabular}{@{}lccl@{}}
\toprule
Cell class & \(c\) & \(\kappa\) & Source \\
\midrule
\emph{E.~coli} run-and-tumble & \(\approx 0.3\) & \(\approx 0.7\) & \cite{BergBrown1972,Berg2004}\\
T cells / lymphocytes in tissue & \(0.4\)--\(0.7\) & \(0.9\)--\(2\) & \cite{MillerEtAl2002,MillerEtAl2003,BeltmanEtAl2007}\\
NK cells in lymph node & \(0.4\)--\(0.7\) & \(0.9\)--\(2\) & \cite{BajenoffEtAl2006}\\
Neutrophils, monocytes in 2D & \(0.5\)--\(0.8\) & \(1\)--\(3\) & \cite{JannatEtAl2010}\\
Fibroblasts, mesenchymal cells & \(0.7\)--\(0.9\) & \(2\)--\(5\) & \cite{GailBoone1970,MaiuriEtAl2015}\\
\emph{Dictyostelium} in vitro & \(0.85\)--\(0.95\) & \(3\)--\(10\) & \cite{VanHaastertBosgraaf2009}\\
Keratocytes & \(\gtrsim 0.95\) & \(\gtrsim 10\) & \cite{EuteneuerSchliwa1984,LeeEtAl1993,KerenEtAl2008}\\
\bottomrule
\end{tabular}
\end{table}

\subsection{T-cell prediction: search-time reduction and diminishing returns}\label{sec:bio-tcell}

We next evaluate the model using parameters representative of T-cell
motion in intact lymph node. We use speed $s = 11~\mu\mathrm{m}/\mathrm{min}$
\cite{MillerEtAl2003}, turning rate $\mu = 0.5~\mathrm{min}^{-1}$ from
persistence time $\tau_p \approx 2~\mathrm{min}$ \cite{BeltmanEtAl2007},
target radius $a = 10~\mu\mathrm{m}$ (an antigen-presenting dendritic
cell with extended dendritic processes) \cite{Lindquist2004}, and
paracortex radius $R = 200~\mu\mathrm{m}$ \cite{BeltmanEtAl2007}.
 The purpose of this calculation is not to reproduce a specific experiment, but to translate the mathematical persistence parameter into a biologically interpretable search time prediction.
\Cref{tab:tcell} reports the angle-averaged MFPT at three representative
starting radii, $r_a = 20~\mu\mathrm{m}$, $r_b = 100~\mu\mathrm{m}$,
and $r_0 = R = 200~\mu\mathrm{m}$, for $\kappa \in \{0, 1, 1.5, 2, 3, 5\}$.
The row $\kappa = 0$ gives the isotropic reorientation baseline, while
$\kappa \approx 1$--$2$ represents the typical lymphocyte range.
Increasing $\kappa$ reduces the search time relative to the baseline,
with most of the improvement concentrated in $\kappa \in [0, 2]$ and
the reduction per unit $\kappa$ falling below 10\% of the
$\kappa = 0 \to 1$ baseline only at $\kappa = 5$. The MFPT therefore
continues to decrease across the validated range, with progressively
diminishing additional benefit.

\begin{table}[t]
  \caption{T cell MFPT prediction across persistence at three starting
    radii, from event-driven Monte Carlo simulation ($M = 10^6$ paths
    per cell, $\pm$~95\% CI shown in parentheses). Biological
    parameters: speed $s = 11~\mu\mathrm{m}/\mathrm{min}$
    \cite{MillerEtAl2003}, turning rate $\mu = 0.5~\mathrm{min}^{-1}$
    from persistence time $\tau_p \approx 2~\mathrm{min}$
    \cite{BeltmanEtAl2007}, target radius $a = 10~\mu\mathrm{m}$
    \cite{Lindquist2004}, paracortex radius $R = 200~\mu\mathrm{m}$
    \cite{BeltmanEtAl2007}. Starting radii are $r_a = 20~\mu\mathrm{m}$
    (near-target probe), $r_b = 100~\mu\mathrm{m}$ (mid-paracortex
    probe), and $r_0 = R = 200~\mu\mathrm{m}$ (outer-paracortex
    probe). MFPT values are in minutes.}
  \label{tab:tcell}
  \centering\small
  \begin{tabular}{@{}cccc@{}}
    \toprule
    $\kappa$ & $\bar T(r_0 = 20~\mu\mathrm{m})$ & $\bar T(r_0 = 100~\mu\mathrm{m})$ & $\bar T(r_0 = 200~\mu\mathrm{m})$ \\
    \midrule
    $0$   & $548.9\ (\pm 1.6)$ & $800.2\ (\pm 1.7)$ & $854.3\ (\pm 1.7)$ \\
    $1$   & $504.2\ (\pm 1.3)$ & $650.2\ (\pm 1.4)$ & $679.9\ (\pm 1.4)$ \\
    $1.5$ & $493.9\ (\pm 1.2)$ & $604.2\ (\pm 1.3)$ & $626.8\ (\pm 1.3)$ \\
    $2$   & $488.7\ (\pm 1.2)$ & $576.9\ (\pm 1.2)$ & $595.1\ (\pm 1.2)$ \\
    $3$   & $485.7\ (\pm 1.1)$ & $554.4\ (\pm 1.1)$ & $566.3\ (\pm 1.1)$ \\
    $5$   & $488.1\ (\pm 1.1)$ & $545.4\ (\pm 1.1)$ & $556.0\ (\pm 1.1)$ \\
    \bottomrule
  \end{tabular}
\end{table}
\begin{table}[t]
  \caption{Applicability of the diffusion limit and boundary-layer
    correction at T-cell parameters. The Milne extrapolation length
    $\ell_{\mathrm{abs}}(\kappa) = s/\lambda(\kappa)$ with
    $\lambda(\kappa) = \mu(1 - m_1(\kappa))$ is not small compared
    with the annulus width $R - a = 190~\mu\mathrm{m}$ at any
    $\kappa$, so the boundary-layer correction does not apply in this parameter regime.}
  \label{tab:applicability}
  \centering\small
  \begin{tabular}{@{}cccc@{}}
    \toprule
    $\kappa$ & $\lambda~(\mathrm{min}^{-1})$ & $\ell_{\mathrm{abs}}~(\mu\mathrm{m})$ & $\ell_{\mathrm{abs}}/(R-a)$ \\
    \midrule
    $0$   & $0.500$ & $22.0$  & $0.116$ \\
    $1$   & $0.277$ & $39.7$  & $0.209$ \\
    $1.5$ & $0.202$ & $54.5$  & $0.287$ \\
    $2$   & $0.151$ & $72.8$  & $0.383$ \\
    $3$   & $0.095$ & $115.8$ & $0.609$ \\
    $5$   & $0.053$ & $206.3$ & $1.086$ \\
    \bottomrule
  \end{tabular}
\end{table}

\begin{figure}[t]
  \centering
  \includegraphics[width=0.55\textwidth]{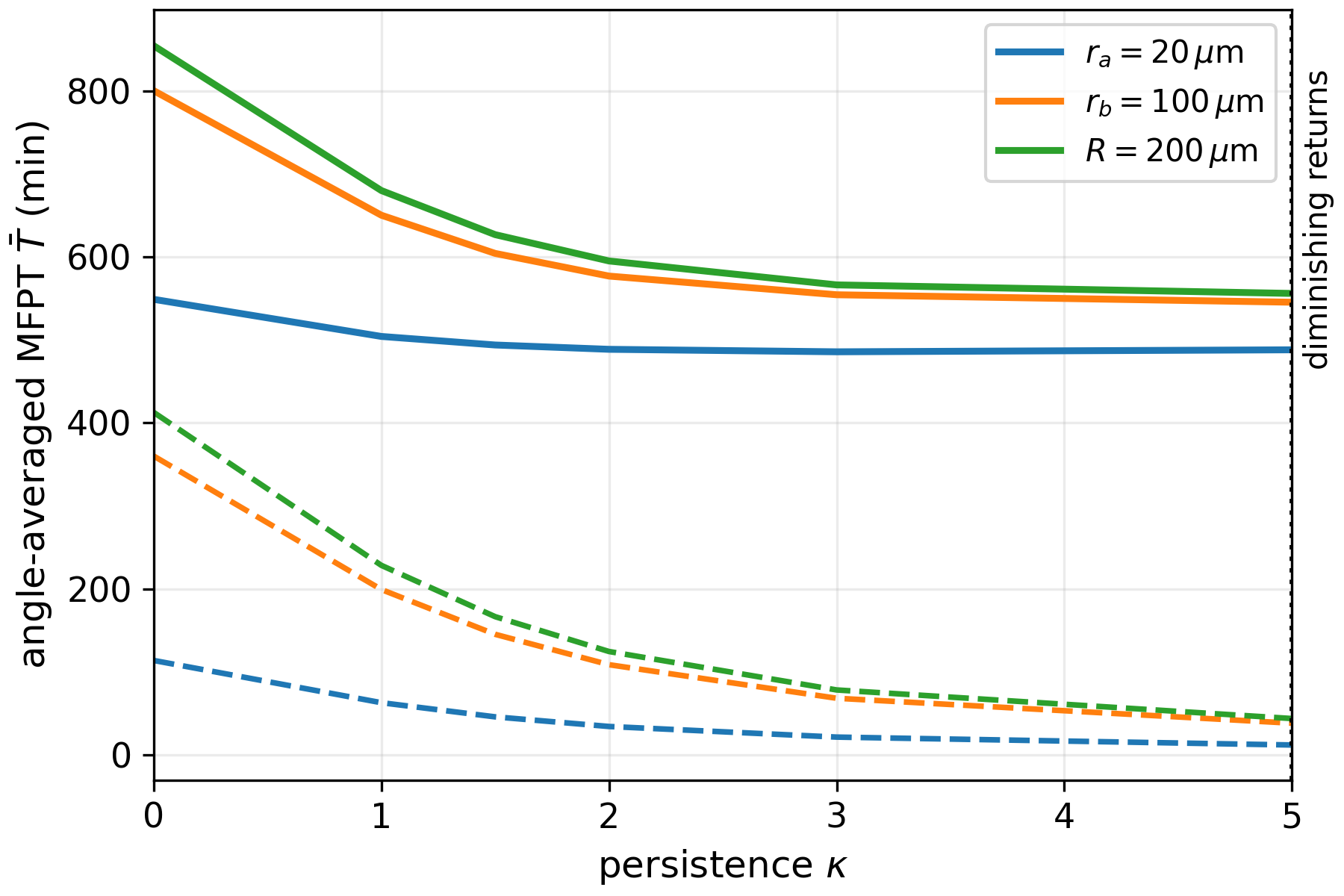}
  \caption{T-cell angle-averaged MFPT $\bar T(r_0; \kappa)$ versus
    persistence parameter $\kappa$ at three starting radii, from
    event-driven Monte Carlo ($M = 10^6$ paths per cell; solid
    curves, $\pm$~CI95 narrower than line width). Dashed curves: 
    diffusion-limit baseline $T_{\mathrm{SRW}}^{D_{\mathrm{eff}}}$ with
    $D_{\mathrm{eff}} = s^2/(2\mu(1 - m_1(\kappa)))$, shown for
    reference. The diffusion limit underestimates the kinetic MFPT
    by a factor of $\sim$4--5 at low $\kappa$ because the mean free
    path $s/\mu = 22~\mu\mathrm{m}$ is comparable to the inner
    geometry $a = 10~\mu\mathrm{m}$ and the diffusion approximation
    does not apply. The boundary layer
    correction is omitted because $\ell_{\mathrm{abs}}(\kappa)/(R - a)$ is not small at any
    tested $\kappa$ (\Cref{tab:applicability}). The vertical dotted line marks the diminishing-returns threshold $\kappa^\star = 5$, at which the marginal reduction in search time
per unit $\kappa$ first falls below 10\% of the $\kappa = 0 \to 1$ baseline.}
  \label{fig:tcell}
\end{figure}

The diffusion-limit baseline lies far below the Monte Carlo MFPT
at all $\kappa$ (\Cref{fig:tcell}). The diffusion
approximation predicts $\sim$115~min at $r_a$ for $\kappa = 0$,
while the kinetic MFPT is $\sim$549~min, a factor of $\sim$5
underestimate. This gap is not a numerical artefact but a
consequence of the parameter regime: with $s/\mu = 22~\mu\mathrm{m}$
comparable to the inner geometry $a = 10~\mu\mathrm{m}$, the
walker frequently turns less than once before reaching the target from the
inner start point, and the fast-turning assumption underlying the
diffusion limit does not hold. The boundary layer correction does not apply either since the extrapolation length $\ell_{\mathrm{abs}}(\kappa)
\sim s/\lambda(\kappa)$ exceeds 10\% of the annulus width $R-a$
already at $\kappa = 0$ and grows with persistence
(\Cref{tab:applicability}). The kinetic transport framework
developed in this paper is therefore not a refinement to the
diffusion limit at T-cell parameters. Rather, it provides the leading-order
prediction itself.
\begin{figure}[h]
  \centering
  \includegraphics[width=0.9\textwidth]{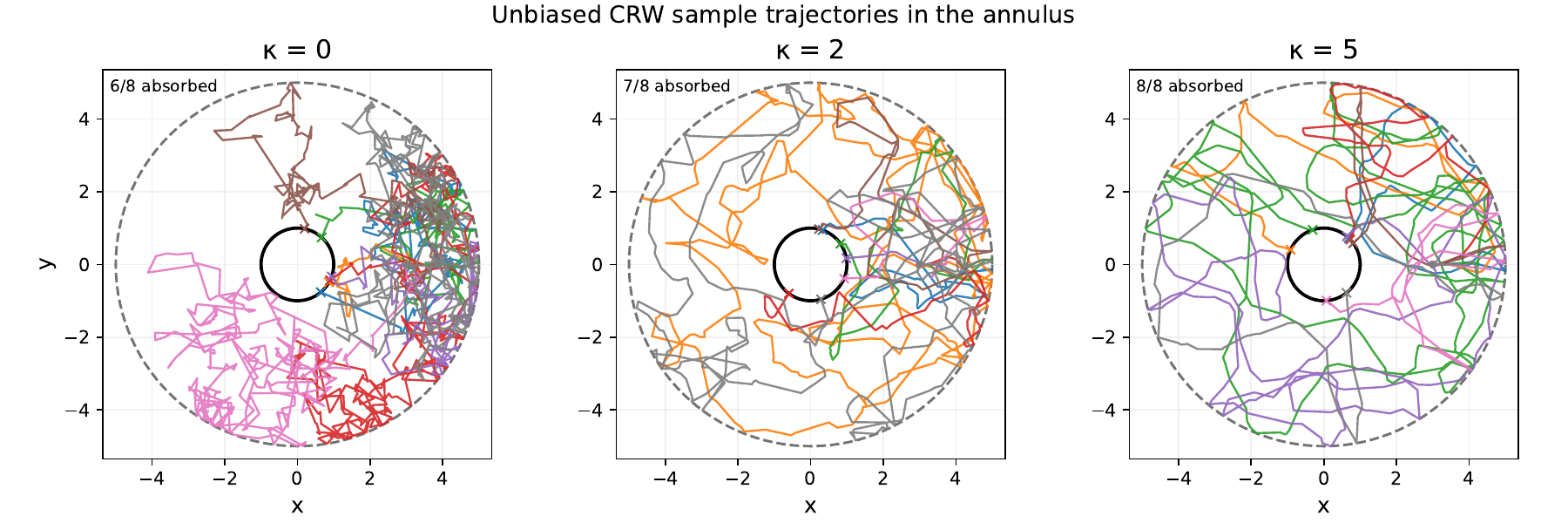}
  \caption{Representative CRW trajectories in the annulus at three persistence levels. Increasing $\kappa$ produces longer straight runs between turns, reduces redundant coverage, and increases the fraction absorbed within the cutoff (shown in each panel).}
  \label{fig:CRWtraj}
\end{figure}
\begin{figure}[h]
  \centering
  \includegraphics[width=0.9\textwidth]{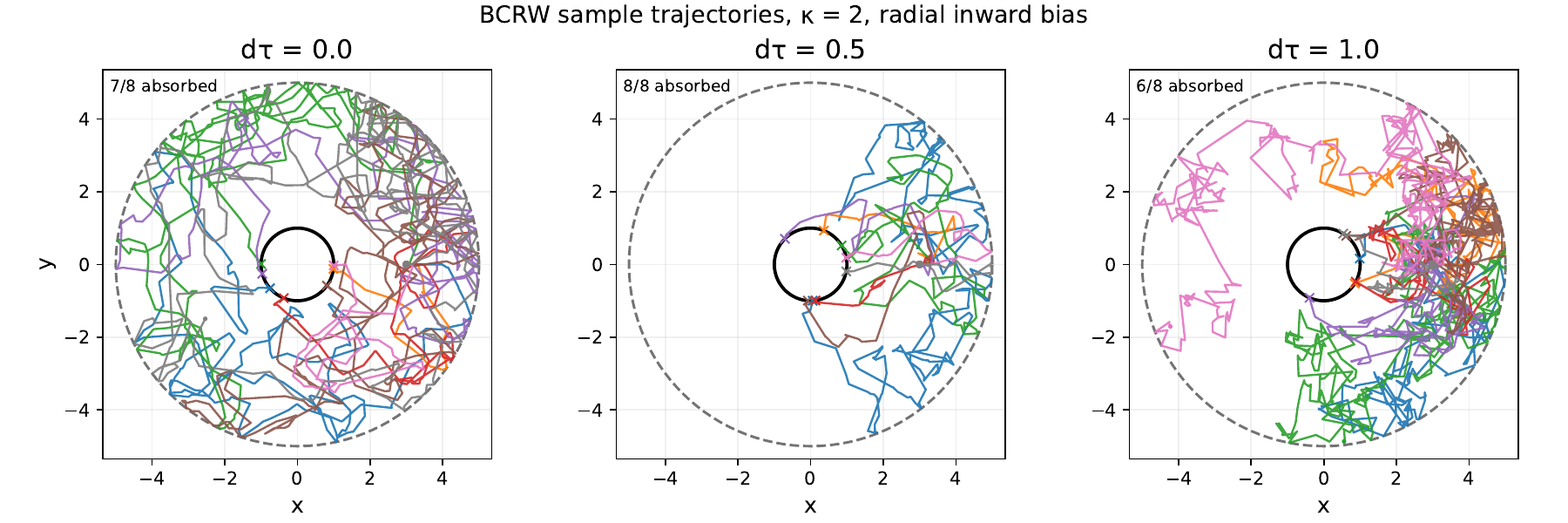}
  \caption{Representative BCRW trajectories at $\kappa=2$ for three bias strengths. Increasing time step length draws paths toward the absorbing target, reducing the exploratory wandering visible at and increasing the absorbed fraction within the cutoff.}
  \label{fig:BCRWtraj}
\end{figure}
%%%%%%%%%%%%%%%%%%%%%%%%%%%%%%%%%%%%%%%%%%%%%%%%%%%%%%%%
\section{Main result}\label{sec:mainresult}

The main conclusion of this study is that directional persistence reduces the mean search time in the annular target problem, but the benefit saturates once persistence is sufficiently large. More precisely, for a correlated random walk in the annulus \(a<r<R\), with absorbing inner boundary at \(r=a\), specular reflection at \(r=R\), constant speed \(s\), turning rate \(\mu\), and von Mises turning kernel with concentration \(\kappa\), the angle-averaged mean first-passage time \(\bar T(r_0;\kappa)\) decreases with \(\kappa\) over the validated range \(\kappa\in[0,5]\). This trend holds at all tested starting radii \(r_0\in(a,R]\), and the reduction becomes weaker near \(\kappa\approx 5\).

This result is supported by several independent calculations. In the fast-turning limit, the correlated random walk reduces to diffusion with effective diffusivity
\[
D_{\rm eff}=\frac{s^2}{2\mu(1-m_1(\kappa))},
\qquad
m_1(\kappa)=\frac{I_1(\kappa)}{I_0(\kappa)}.
\]
Since \(m_1(\kappa)\) increases with \(\kappa\), the effective diffusivity also increases with \(\kappa\). The diffusion-limit MFPT therefore decreases as persistence increases. The absorbing-boundary correction derived in \Cref{app:eps-correction} preserves this interpretation: finite persistence shifts the MFPT by a boundary-layer offset rather than changing the leading radial dependence in the bulk.

The numerical results support the same conclusion beyond the
diffusion limit. The diffusion approximation accounts for the
leading behaviour at moderate persistence, but as $\kappa$ grows
the walker enters the strong-persistence regime described in
the last part of \Cref{sec:analysis}: the von Mises kernel concentrates
near zero turning angle, the angular turning operator is
approximated by angular diffusion with $D_\psi = \mu/\kappa$, and
the radial motion becomes nearly deterministic between rare
changes in heading. In this limit the MFPT approaches the
ballistic-hitting time set by the chord geometry of the annulus
rather than the diffusion-limit MFPT, so $\bar T(r_0; \kappa)$
continues to decrease with $\kappa$ but with diminishing additional
benefit. The two analytical limits therefore bound the validated
range from below and above, both predicting a monotone reduction
in search time with persistence.

The semi-Lagrangian $(r, \psi)$ solver reproduces this monotone
decrease of $\bar T(r_0; \kappa)$ at all probe radii
$r_0 \in \{1.5, 2.5, 3.5, 5\}$ over the validated range
$\kappa \in [0, 5]$. The event-driven Monte Carlo simulator gives an
independent check of the same trend, including the high-persistence
comparison at $r = 5$. The time step-free stationary formulation
further confirms the trend at $\kappa = 5$ without the structural
accuracy ceiling associated with the pseudo-time semi-Lagrangian
formulation. Together, these analytical, deterministic numerical,
and Monte Carlo results show that persistence accelerates confined
target search over the biologically relevant range, with diminishing
additional benefit as the motion approaches the strong-persistence
regime.

\section{Discussion}\label{sec:discussion}

The monotone reduction of search time with increasing persistence is consistent with the broader random-walk literature in biology and ecology, where directional persistence often reduces encounter times by allowing a walker to explore space more efficiently \cite{CodlingPlankBenhamou2008,CodlingThesis2003}. The contribution of the present work is to quantify this effect for a confined annular search geometry with an absorbing central target and a specularly reflecting outer boundary. In this setting, persistence improves search efficiency over the biologically relevant range \(\kappa\in[0,5]\), but the improvement begins to saturate near the upper end of this range.

The analysis also clarifies how the correlated random walk connects to simpler diffusion-based models. In the fast-turning regime, persistence enters through the effective diffusivity \(D_{\rm eff}\), giving a direct analytical explanation for why increasing \(\kappa\) reduces the MFPT. The next correction is not a smooth \(r\)-dependent bulk correction in the annulus. Instead, the leading finite-persistence effect appears as an absorbing-boundary, Milne-type offset. This provides a simple way to estimate the deviation from the diffusion-limit prediction without solving the full transport problem.

The numerical workflow separates the regimes in which different approximations are reliable. For \(\kappa\in[0,5]\), the semi-Lagrangian solver, the stationary formulation, and Monte Carlo simulations give consistent trends. For larger \(\kappa\), the motion enters a strong-persistence regime in which the near-wall layer becomes difficult to resolve on accessible semi-Lagrangian grids. In that regime, Monte Carlo and the strong-persistence analysis become the more reliable references. This is also the regime relevant to highly persistent cells such as keratocytes and some \emph{Dictyostelium} conditions.

Several limitations remain. The present model is restricted to a two-dimensional, rotationally symmetric annulus with a single absorbing target at the center. It does not include three-dimensional tissue geometry, multiple targets, asymmetric confinement, cell-cell interactions, spatially varying speeds, or Lévy-walk behavior \cite{HarrisEtAl2012}. The biased correlated random walk formulation developed here also remains to be completed numerically for nontrivial chemotactic fields.

Natural extensions include applying the time step-free stationary formulation as the main production solver, completing the BCRW solver for spatially varying guidance cues, extending the boundary-layer correction beyond leading order, and moving from the annulus to three-dimensional spheroid and more complex multi-target geometries. The recent higher-dimensional velocity-jump MFPT framework of~\cite{d2026mean}, which develops a Langevin approximation and explicit MFPT formulae for von Mises and Fisher kernels in $d = 2, 3$, provides a natural template for the three-dimensional extension and for incorporating chemotactic bias through a spatially varying preferred direction.% $\hat\gamma(\mathbf{x})$.
%%%%%%%%%%%%%%%%
\newpage
\appendix
\section{Finite-difference benchmarks for SRW and
BRW}\label{app:fd}
This appendix explains the second order centered finite difference solver used to verify the analytical hierarchy of \Cref{sec:analysis-srwbrw}. The radial interval $[a,R]$ is discretized
uniformly with $h=(R-a)/(N_r-1)$. At each interior node the differential operator is replaced by its centered difference approximation, and the reflecting outer boundary by the second order backward difference $3u_{N_{r-1}}-4u_{N_{r-2}}+u_{N_{r-3}}=0$.

\Cref{tab:srw-conv} reports the constant diffusivity SRW
grid convergence study at $a=1$, $R=5$, $D=1$; the observed order approaches~$2$. The observed second-order convergence confirms the finite-difference implementation; the same scheme is applied without modification to the variable-coefficient SRW and BRW cases used as simulations in \Cref{sec:analysis}.

\begin{table}[h]
\caption{Constant-diffusivity SRW grid-convergence study, $a=1$, $R=5$,
$D=1$.}
\label{tab:srw-conv}
\centering\small
\begin{tabular}{@{}rcccc@{}}
\toprule
$N_r$ & $h$ & $\max_i|u_\text{num}-u_\text{exact}|$ & max rel.\ err & order\\
\midrule
$50$  & $8.16\times10^{-2}$ & $8.99\times10^{-4}$ & $3.20\times10^{-4}$ & --\\
$100$ & $4.04\times10^{-2}$ & $2.26\times10^{-4}$ & $8.45\times10^{-5}$ & $1.96$\\
$200$ & $2.01\times10^{-2}$ & $5.66\times10^{-5}$ & $2.17\times10^{-5}$ & $1.98$\\
$400$ & $1.00\times10^{-2}$ & $1.42\times10^{-5}$ & $5.49\times10^{-6}$ & $1.99$\\
\bottomrule
\end{tabular}
\end{table}
%%%%%%%%%%%%%%%%%%%%%%%%%%%%%%%%%%%%%%%%%%%%%%%%%%%%%%%%%%%%

\section{Specular reflection at the outer boundary}\label{app:specular}
This appendix explains the derivation of the specular reflection condition and places it in the general framework of kinetic boundary conditions for transport equations.

\begin{figure}[h]
\centering
\begin{tikzpicture}[x=0.75pt,y=0.75pt,yscale=-0.7,xscale=0.7]
	%uncomment if require: 
%	\path (50,150); %set diagram left start at 0, and has height of 300
	
	%Shape: Arc [id:dp09127818026341905] 
	\draw  [draw opacity=0] (425.42,88.74) .. controls (493.77,88.67) and (558.26,124.44) .. (590.77,185.75) -- (420.08,275.25) -- cycle ; \draw   (425.42,88.74) .. controls (493.77,88.67) and (558.26,124.44) .. (590.77,185.75) ;  
	%Shape: Axis 2D [id:dp24526464624083488] 
	\draw  (423.14,249.82) -- (495.65,249.82)(430.4,190.06) -- (430.4,256.46) (488.65,244.82) -- (495.65,249.82) -- (488.65,254.82) (425.4,197.06) -- (430.4,190.06) -- (435.4,197.06)  ;
	%Shape: Axis 2D [id:dp4332303210737607] 
	\draw [color={rgb, 255:red, 41; green, 30; blue, 197 }  ,draw opacity=1 ] (510.19,107.31) -- (582.27,152.43)(554.26,52.92) -- (513.3,118.37) (578.99,144.47) -- (582.27,152.43) -- (573.68,152.95) (546.31,56.2) -- (554.26,52.92) -- (554.78,61.51)  ;
	%Shape: Axis 2D [id:dp38339659293506023] 
	\draw [color={rgb, 255:red, 41; green, 30; blue, 197 }  ,draw opacity=1 ] (423.19,245.31) -- (495.27,290.43)(467.26,190.92) -- (426.3,256.37) (491.99,282.47) -- (495.27,290.43) -- (486.68,290.95) (459.31,194.2) -- (467.26,190.92) -- (467.78,199.51)  ;
	%Straight Lines [id:da07309370859472408] 
	\draw [color={rgb, 255:red, 41; green, 30; blue, 197 }  ,draw opacity=1 ] [dash pattern={on 0.84pt off 2.51pt}]  (517.4,111.82) -- (467.14,191.52) ;
	%Shape: Arc [id:dp5357616028067176] 
	\draw  [draw opacity=0] (441.08,232.06) .. controls (447.86,236.01) and (452.95,242.55) .. (455.01,250.31) -- (426,258) -- cycle ; \draw  [color={rgb, 255:red, 41; green, 30; blue, 197 }  ,draw opacity=1 ] (441.08,232.06) .. controls (447.86,236.01) and (452.95,242.55) .. (455.01,250.31) ;  
	%Straight Lines [id:da14553777611609264] 
	\draw [line width=1.5]    (404.14,130.52) -- (514.44,112.31) ;
	\draw [shift={(517.4,111.82)}, rotate = 170.63] [color={rgb, 255:red, 0; green, 0; blue, 0 }  ][line width=1.5]    (14.21,-4.28) .. controls (9.04,-1.82) and (4.3,-0.39) .. (0,0) .. controls (4.3,0.39) and (9.04,1.82) .. (14.21,4.28)   ;
	%Straight Lines [id:da3842163261745698] 
	\draw [color={rgb, 255:red, 209; green, 29; blue, 51 }  ,draw opacity=1 ][line width=1.5]    (582.51,213) -- (517.4,111.82) ;
	\draw [shift={(584.14,215.52)}, rotate = 237.23] [color={rgb, 255:red, 209; green, 29; blue, 51 }  ,draw opacity=1 ][line width=1.5]    (14.21,-6.37) .. controls (9.04,-2.99) and (4.3,-0.87) .. (0,0) .. controls (4.3,0.87) and (9.04,2.99) .. (14.21,6.37)   ;
	%Straight Lines [id:da27337813474289296] 
	\draw    (399.14,131.52) -- (500.14,131.52) ;
	%Straight Lines [id:da39366133686582394] 
	\draw    (517.4,111.82) -- (618.4,111.82) ;
	%Shape: Arc [id:dp5934870273616739] 
	\draw  [draw opacity=0] (451.08,121.65) .. controls (453.66,124.2) and (455.78,127.22) .. (457.31,130.57) -- (430,143) -- cycle ; \draw  [color={rgb, 255:red, 0; green, 0; blue, 0 }  ,draw opacity=1 ] (451.08,121.65) .. controls (453.66,124.2) and (455.78,127.22) .. (457.31,130.57) ;  
	%Shape: Arc [id:dp7372976483751559] 
	\draw  [draw opacity=0] (541.75,113.18) .. controls (542.5,115.54) and (542.9,118) .. (542.9,120.52) .. controls (542.9,127.1) and (540.2,133.22) .. (535.57,138.31) -- (499.45,120.52) -- cycle ; \draw  [color={rgb, 255:red, 209; green, 29; blue, 51 }  ,draw opacity=1 ] (541.75,113.18) .. controls (542.5,115.54) and (542.9,118) .. (542.9,120.52) .. controls (542.9,127.1) and (540.2,133.22) .. (535.57,138.31) ;  
	
	% Text Node
	\draw (425,169.4) node [anchor=north west][inner sep=0.75pt]  [font=\footnotesize]  {$y$};
	% Text Node
	\draw (501,242.4) node [anchor=north west][inner sep=0.75pt]  [font=\footnotesize]  {$x$};
	% Text Node
	\draw (560,39.4) node [anchor=north west][inner sep=0.75pt]  [font=\small,color={rgb, 255:red, 41; green, 30; blue, 197 }  ,opacity=1 ]  {$\overrightarrow{e_{r}}$};
	% Text Node
	\draw (590,141.4) node [anchor=north west][inner sep=0.75pt]  [font=\small,color={rgb, 255:red, 41; green, 30; blue, 197 }  ,opacity=1 ]  {$\overrightarrow{e_{\theta }}$};
	% Text Node
	\draw (454,229.4) node [anchor=north west][inner sep=0.75pt]  [font=\small,color={rgb, 255:red, 41; green, 30; blue, 197 }  ,opacity=1 ]  {$\theta $};
	% Text Node
	\draw (491,75.4) node [anchor=north west][inner sep=0.75pt]    {$\vec{V}_{\text{old}}$};
	% Text Node
	\draw (586.14,218.92) node [anchor=north west][inner sep=0.75pt]  [color={rgb, 255:red, 209; green, 29; blue, 51 }  ,opacity=1 ]  {$\vec{V}_{\text{new}}$};
	% Text Node
	\draw (428,105.4) node [anchor=north west][inner sep=0.75pt]  [font=\small]  {$\phi '$};
	% Text Node
	\draw (551,114.4) node [anchor=north west][inner sep=0.75pt]  [font=\small,color={rgb, 255:red, 209; green, 29; blue, 51 }  ,opacity=1 ]  {$\phi $};
	
\end{tikzpicture}
\caption{Specular reflection in the local frame at the outer
boundary $r=R$.\label{fig:specular}}
\end{figure}
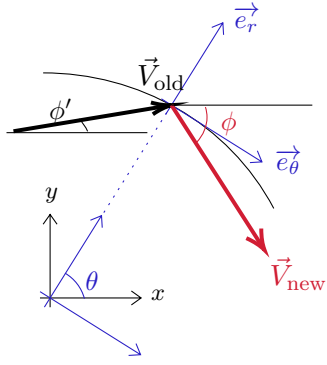

At the outer wall the local frame is
$\vec e_r(\theta) = (\cos\theta,\sin\theta)$,
$\vec e_\theta(\theta) = (-\sin\theta,\cos\theta)$.
The incoming velocity $\vec V_{\text{old}} = s(\cos\phi,\sin\phi)$
projects onto this frame as
\begin{equation}
v_r = \vec V_{\text{old}}\cdot\vec e_r = s\cos(\phi-\theta),\qquad
v_\theta = \vec V_{\text{old}}\cdot\vec e_\theta = s\sin(\phi-\theta),
\end{equation}
so that
$\vec V_{\text{old}} = s\cos(\phi-\theta)\,\vec e_r
                     + s\sin(\phi-\theta)\,\vec e_\theta$.
Specular reflection flips the radial component and preserves the
tangential component,
\begin{equation}
\vec V_{\text{new}} = -v_r\,\vec e_r + v_\theta\,\vec e_\theta
                    = -s\cos(\phi-\theta)\,\vec e_r
                    + s\sin(\phi-\theta)\,\vec e_\theta.
\end{equation}
Writing this in angle space, with $\phi'$ the incoming direction and
$\phi$ the reflected direction,
$\phi' - \theta \mapsto \pi - (\phi - \theta)$, so that
\begin{equation}
\phi(\theta,\phi') = 2\theta + \pi - \phi'.
\label{eq:spec-angle}
\end{equation}
In the heading variable $\psi = \phi - \theta$, which is the natural
variable for the rotationally symmetric unbiased problem,
\Cref{eq:spec-angle} reduces to the CRW boundary condition
\begin{equation}
T(R,\psi) = T(R,\pi-\psi).
\label{eq:spec-psi}
\end{equation}
For the BCRW, the same identity
\eqref{eq:bcrw-reflect} is kept in the general form. Schematics is shown in \Cref{fig:specular}.

%%%%%%%%%%%%%%%%%%%%%%%%%%%%%%%%%%%%%%%%%%%%%%%%%%%%%%
\section{BCRW Fourier hierarchy}\label{app:bcrw}

This appendix focuses on the full mode-space form of the biased correlated random walk (BCRW) backward equation \eqref{eq:bcrw-bte}. We expand the MFPT in the two angular variables as
\[
T(r,\theta,\phi)
=
\sum_{m,n\in\mathbb{Z}}
U_{m,n}(r)\mathrm{e}^{im\theta}\mathrm{e}^{in\phi}.
\]
in the streaming terms, then couple neighboring angular modes. In particular,
\begin{align}
[\cos(\phi-\theta)\partial_rT]_{m,n}
  &=
  \frac12\bigl(U_{m+1,n-1}'(r)+U_{m-1,n+1}'(r)\bigr),\\
\left[\frac{\sin(\phi-\theta)}{r}\partial_\theta T\right]_{m,n}
  &=
  \frac{1}{2r}
  \bigl((m+1)U_{m+1,n-1}(r)-(m-1)U_{m-1,n+1}(r)\bigr).
\end{align}
Combining these identities with the Fourier representation of the turning operator in \eqref{eq:bcrw-fourier-turn}, and matching coefficients of
$\mathrm{e}^{im\theta}\mathrm{e}^{in\phi}$, gives the coupled radial hierarchy
\begin{multline}
-\delta_{m,0}\delta_{n,0}
=
\frac{s}{2}\bigl(U'_{m+1,n-1}+U'_{m-1,n+1}\bigr)
+
\frac{s}{2r}\bigl((m+1)U_{m+1,n-1}-(m-1)U_{m-1,n+1}\bigr)\\
+\mu\left[
\sum_{q\in\mathbb Z}
\frac{I_n(\kappa)}{I_0(\kappa)}
C_{n,q}\,U_{m-q,n}(r)
-
U_{m,n}(r)
\right].
\label{eq:bcrw-mode-system}
\end{multline}
The absorbing boundary condition gives
\[
U_{m,n}(a)=0
\qquad
\text{for all } m,n,
\]
while specular reflection at the outer boundary gives
\[
U_{m,n}(R)=(-1)^n U_{m+2n,-n}(R).
\]
Here $C_{n,q}$ denotes the $q$th Fourier coefficient, in the variable $\theta$, of
$\mathrm{e}^{in\delta(r,\theta)}$, where $\delta(r,\theta)$ is the phase shift associated with the bias direction. Thus the biased turning operator is no longer diagonal in the position mode $m$; instead, the spatial dependence of the bias introduces a convolution over $m$.

In the unbiased limit $\delta\equiv 0$, the coefficients satisfy
\[
C_{n,0}=1,
\qquad
C_{n,q}=0 \quad \text{for } q\neq 0.
\]
The convolution then collapses, and the turning operator reduces to the diagonal CRW operator in \eqref{eq:K-fourier-crw}. This recovers the unbiased CRW hierarchy \eqref{eq:crw-fourier-hierarchy}.

%%%%%%%%%%%%%%%%%%%%%%%%%%%%%%%%%%%%%%%%%%%%%%%%%%%%%%%
\section{Compatibility issues in truncated mode-space formulations}\label{app:dae}

Before adopting the direct $(r,\psi)$ transport solver described in \Cref{sec:numerics}, we tested a family of truncated-Fourier formulations for \Cref{eq:crw-backward}. These formulations are recorded here to clarify why they were not used as the main numerical method.

After Fourier reduction and truncation, the resulting radial system can be written in the form
\[
M w'(r)+B(r)w(r)=f,
\]
where the matrix $M$ is singular. Consequently, any left-null vector $\ell$ of $M$ imposes an algebraic compatibility condition,
\[
\ell^\top B(r)w(r)=\ell^\top f .
\]
In practice, this compatibility condition was not simultaneously consistent with the absorbing boundary data and the finite-mode truncation, even in the lowest nontrivial truncation. We observed the same issue across several variants, including buffered differential algebraic systems, retained mode reductions with $N=1$ and $N=2$, asymptotic closures for omitted modes, and finite-$K$ non-buffered spectral reference systems.

In each case, the transformed discrete system could be solved to high numerical accuracy, but the residual in the original physical transport equation did not decrease under refinement, and in some cases increased. Block elimination, Schur-complement methods, direct solvers, and QR-based consistency solvers all led to the same physical residual behavior. This indicates that the main issue is the finite-mode formulation itself rather than the choice of linear algebra solver. For this reason, the mode-space systems are used only as analytical and diagnostic tools in the present work, while the primary computations are based on the direct stationary transport discretization.
%%%%%%%%%%%%%%%%%%%%%%%%%%%%%%%%%%%%%%%%%%%%%%%%%%%%%%%%%%%%%%%%%
\section{Boundary-layer correction in the annulus}
\label{app:eps-correction}

We summarize the calculation leading to the absorbing-boundary correction used in
Subsection 3.1. Let
\[
    \lambda=\mu(1-m_1), \qquad 
    m_1=\frac{I_1(\kappa)}{I_0(\kappa)}, \qquad
    D_{\mathrm{eff}}=\frac{s^2}{2\lambda}.
\]
The leading outer solution is the diffusion-limit MFPT
\[
    D_{\mathrm{eff}}\Delta T_0=-1,\qquad
    T_0(a)=0,\qquad \partial_rT_0(R)=0,
\]
and therefore
\begin{equation}
       T_0(r)=\frac{R^2}{2D_{\mathrm{eff}}}\log\frac{r}{a}
    +\frac{a^2-r^2}{4D_{\mathrm{eff}}}.
\end{equation}

We write the transport MFPT as
\[
    T_{\mathrm{CRW}}(r;\kappa)=T_0(r)+\frac{1}{\lambda}T_1(r)+\cdots .
\]
The Chapman--Enskog correction to the bulk equation \cite{HillenOthmer2000,OthmerHillen2002} gives
\[
    D_{\mathrm{eff}}\Delta T_1=D_{\mathrm{eff}}^2\Delta^2T_0 .
\]
Since \(D_{\mathrm{eff}}\Delta T_0=-1\) in the annulus, \(\Delta^2T_0=0\), and hence
\[
    \Delta T_1=0 .
\]
For a radial correction this gives
\[
    T_1(r)=A\log r+B .
\]
The outer boundary is specularly reflecting, so no absorbing kinetic layer is generated at
\(r=R\), and the outer expansion inherits the reflecting condition order by order:
\[
    \partial_rT_1(R)=0 .
\]
Thus \(A=0\), and the first regular correction is a constant independent of the starting
radius.

The value of this constant is set by the absorbing boundary layer at \(r=a\). The transport
condition is a half-range condition, imposed only on headings entering the target, rather
than the scalar diffusion condition \(T(a)=0\). In the outer diffusion problem this is replaced
by the Milne-type extrapolation condition
\[
    T_{\mathrm{out}}(a)=\ell_{\mathrm{abs}}\partial_rT_{\mathrm{out}}(a),
    \qquad
    \ell_{\mathrm{abs}}=\chi_{\mathrm{abs}}(\kappa)\frac{s}{\lambda},
\]
where \(\chi_{\mathrm{abs}}(\kappa)\) is the dimensionless half-space extrapolation constant
for the von Mises turning kernel.

At this order \(T_{\mathrm{out}}=T_0+B\), and since \(T_0(a)=0\),
\[
    B=\ell_{\mathrm{abs}}T_0'(a).
\]
Using
\[
    T_0'(a)=\frac{R^2-a^2}{2D_{\mathrm{eff}}a},
    \qquad
    D_{\mathrm{eff}}=\frac{s^2}{2\lambda},
\]
we obtain
\[
    B=\chi_{\mathrm{abs}}(\kappa)\frac{R^2-a^2}{as}.
\]
Therefore
\begin{equation}
     T_{\mathrm{CRW}}(r;\kappa)
    =
    T_{\mathrm{SRW}}^{D_{\mathrm{eff}}}(r)
    +
    \chi_{\mathrm{abs}}(\kappa)\frac{R^2-a^2}{as}
    +\text{higher-order terms}.
\end{equation}
The correction is independent of \(r\) because the regular interior correction has a zero Laplacian and the reflecting boundary removes the logarithmic mode. Its magnitude is set by the absorbing kinetic layer through \(\chi_{\mathrm{abs}}(\kappa)\).

The constant \(\chi_{\mathrm{abs}}(\kappa)\) is a Milne-type extrapolation constant: it is the dimensionless constant obtained by solving the half-space kinetic boundary-layer problem associated with the half-range absorbing boundary condition. Such extrapolation constants are standard in diffusion-limit analyses of kinetic and transport equations, where the kinetic boundary layer converts a microscopic inflow or absorbing boundary condition into an effective boundary condition for the outer diffusion problem \cite{Cercignani1969,Klar2004,Guo2025Transport}.

This result also gives two useful consistency checks. When \(\kappa=0\), the kernel is isotropic and \(\lambda=\mu\). The extrapolation constant \(\chi_{\mathrm{abs}}(0)\) is finite, so the calculation predicts a nonzero offset between the diffusion-limit SRW formula and the full velocity-jump MFPT even for isotropic reorientation. This offset is not a solver error; it is the absorbing-wall kinetic correction. In contrast, as \(\kappa\to\infty\), \(m_1\to 1\), so \(\lambda\to 0\) and the extrapolation length \(s/\lambda\) becomes large. The Milne layer is then no longer thin relative to the annulus, and the diffusion expansion breaks down. This is the strong persistence regime treated separately in the numerical verification.

% \begin{reading}
% \end{reading}

\section*{Acknowledgments}
We thank Thomas Hillen, Alan Lindsay, Sarafa Iyaniwura and especially Michael Ward for helpful discussions during the development of this project.
\section*{Author Contributions}
FS: Problem development; model set-up; analytical work and approximations; numerical methods and testing; writing the manuscript. DC: Problem development; analytical work and approximations; writing the manuscript.
\section*{Access to Code}
At present, please send requests for code to the authors.
\bibliographystyle{siamplain}
\bibliography{references}

\pagebreak

{\bf Supplementary Figures}

\begin{figure}[h]
  \centering
  \includegraphics[width=0.95\textwidth]{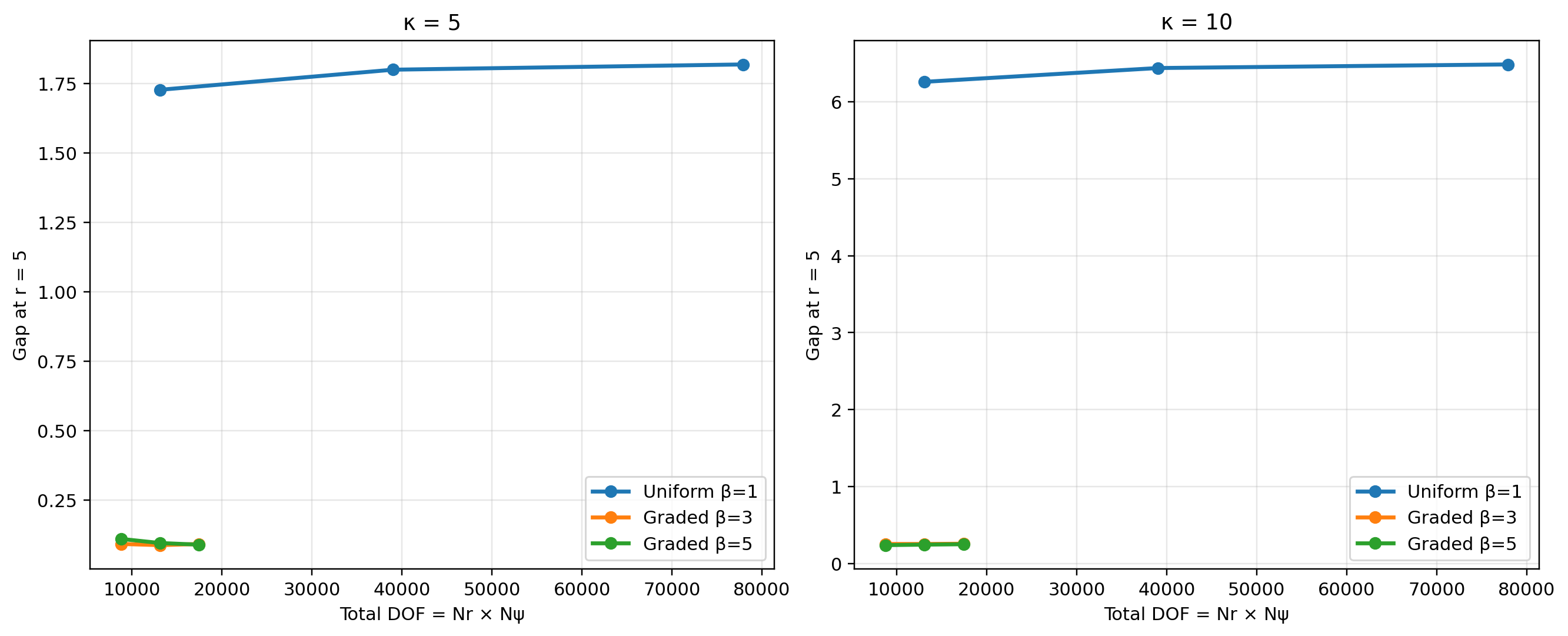}
  \caption{Direct-vs-Monte-Carlo relative gap at $r = 5$ on uniform and
    power-law graded radial meshes, as a function of total angular plus
    radial degrees of freedom $N_r \times N_\psi$. 
    These two values bracket the regime boundary: $\kappa = 5$ is the
  upper end of the biologically relevant range and the hardest case the production solver must resolve, while
  $\kappa = 10$ lies in the strong-persistence regime where the wall
  layer falls below the minimum graded mesh cell width.
    Three mesh families are compared at
    $N_\psi = 144$, $\Delta\tau = 0.007$: uniform ($\beta = 1$), graded
    $\beta = 3$, and graded $\beta = 5$. At $\kappa = 5$, graded
    $\beta = 3$ collapses the near-$R$ wall error by more than an order
    of magnitude at fixed total DOF and is essentially flat under further
    refinement, justifying its choice as the production setting; graded
    $\beta = 5$ gives no additional benefit and drives $\Delta r_{\min}$
    toward machine epsilon.}
  \label{fig:SI-beta-sweep}
\end{figure}

\begin{figure}[htbp]
  \centering
  \includegraphics[width=0.75\textwidth]{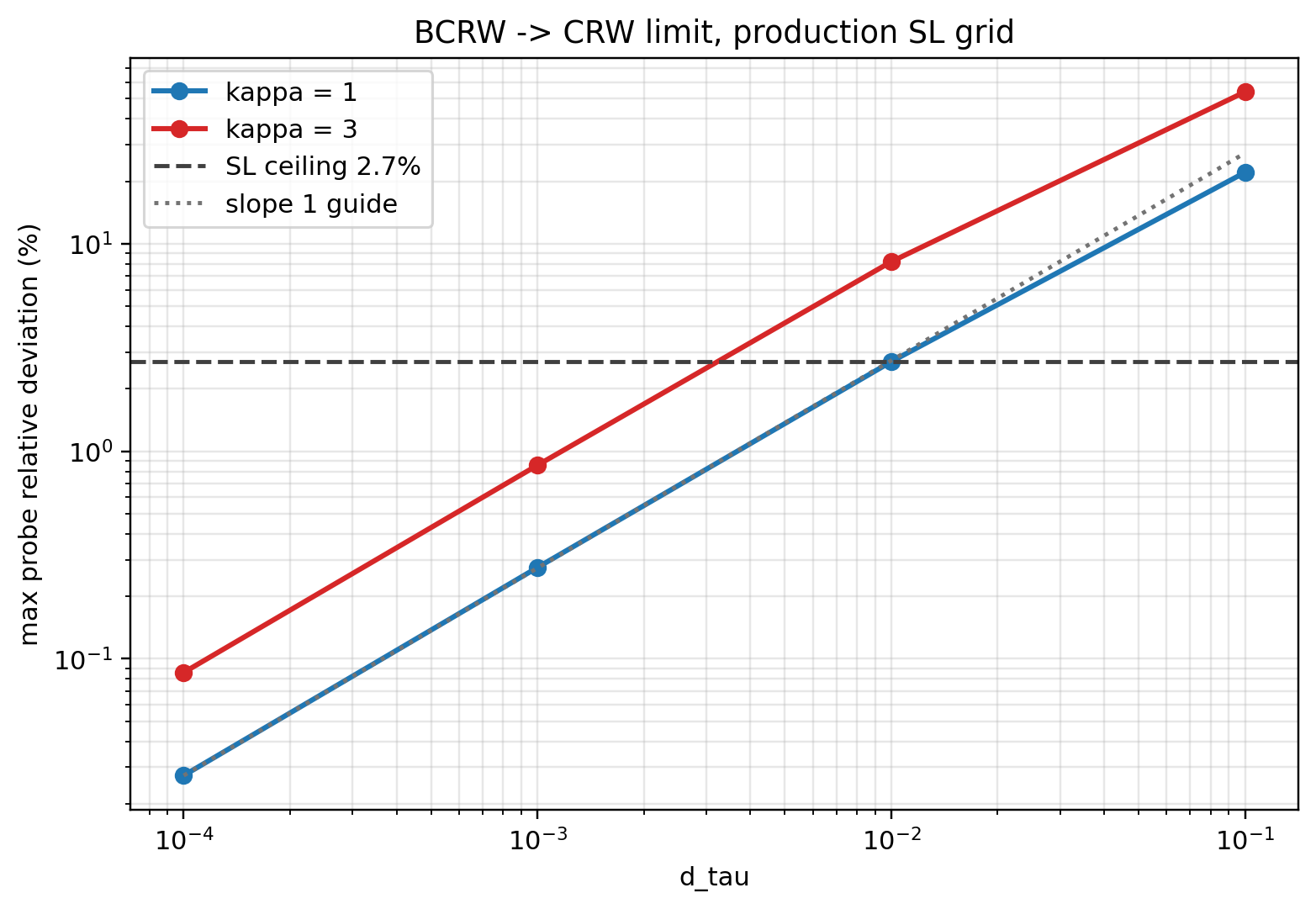}
  \caption{BCRW-to-CRW consistency check at the production
    semi-Lagrangian grid ($\beta = 3$, $N_r = 121$, $N_\psi = 144$,
    $a = 1$, $R = 5$, $s = 1$, $\mu = 3$). Plotted is the maximum probe
    relative deviation
    $|\bar T_{\mathrm{BCRW}} - \bar T_{\mathrm{CRW}}| /
    \bar T_{\mathrm{CRW}}$ over the four probe radii
    $r \in \{1.5, 2.5, 3.5, 5\}$, as a function of bias strength
    $d_\tau$, for $\kappa = 1$ (lymphocyte range) and $\kappa = 3$
    (fibroblast range). The dashed line marks the semi-Lagrangian
    structural ceiling at $2.7\%$; the dotted line is a slope-1
    reference. Fitted slopes are $0.997$ and $0.990$ respectively,
    confirming the $O(d_\tau)$ scaling expected from the kernel
    expansion. The crossover with the structural ceiling near
    $d_\tau \approx 10^{-2}$ sets the bias range over which the BCRW
    and unbiased CRW solvers can be compared directly.}
  \label{fig:SI-bcrw-dtau-limit}
\end{figure}

\end{document}